\documentclass[11pt,a4paper]{article}
\usepackage[T1]{fontenc}
\usepackage[utf8]{inputenc}
\usepackage{lmodern}
\usepackage[margin=1in]{geometry}
\usepackage{microtype,amsmath,amssymb,booktabs,array,longtable,tabularx}
\usepackage{graphicx,xcolor,caption,enumitem,placeins}
\graphicspath{{./}}
\usepackage[hidelinks,unicode]{hyperref}
\usepackage{fancyhdr}
\setlist{nosep,leftmargin=*}
\newcommand{\M}[2]{\mbox{\csname m@#1@#2\endcsname\%}}
\newcommand{\E}[4]{\csname e@#1@#2@#3@#4\endcsname}
\newcommand{\WN}{\texttt{warm\_norm}}
\newcommand{\WU}{\texttt{warm\_unit}}
\newcommand{\NA}{\textemdash}
\newcommand{\rr}{\textsuperscript{r}}
\newcommand{\pp}{\textsuperscript{p}}
\newcommand{\ssm}{\textsuperscript{s}}
\newcommand{\smallnote}[1]{\par\vspace{3pt}{\small #1\par}}
\hypersetup{pdftitle={Cross-Day Compatibility of Daily Item Embeddings: An Empirical Study of Alignment, Scale Management, and CTR Adaptation},pdfauthor={Yicheng Wang},pdfsubject={Daily co-occurrence embeddings and multi-version CTR learning}}
\expandafter\def\csname m@scratch_raw@cosine\endcsname{0.0}
\expandafter\def\csname m@scratch_raw@l2\endcsname{0.0}
\expandafter\def\csname m@scratch_raw@norm\endcsname{0.0}
\expandafter\def\csname m@scratch_raw@recall\endcsname{0.0}
\expandafter\def\csname m@scratch_raw@gauc\endcsname{0.0}
\expandafter\def\csname m@scratch_raw@logloss\endcsname{0.0}
\expandafter\def\csname m@scratch_raw@p2v_minutes\endcsname{0.0}
\expandafter\def\csname m@scratch_raw@dnn_minutes\endcsname{0.0}
\expandafter\def\csname m@xsimgcl_raw@cosine\endcsname{-102.4}
\expandafter\def\csname m@xsimgcl_raw@l2\endcsname{+31.2}
\expandafter\def\csname m@xsimgcl_raw@norm\endcsname{-0.9}
\expandafter\def\csname m@xsimgcl_raw@recall\endcsname{+76.4}
\expandafter\def\csname m@xsimgcl_raw@gauc\endcsname{-0.5}
\expandafter\def\csname m@xsimgcl_raw@logloss\endcsname{+0.2}
\expandafter\def\csname m@xsimgcl_raw@p2v_minutes\endcsname{+22.2}
\expandafter\def\csname m@xsimgcl_raw@dnn_minutes\endcsname{\NA}
\expandafter\def\csname m@scratch_procrustes@cosine\endcsname{+93.4}
\expandafter\def\csname m@scratch_procrustes@l2\endcsname{-48.2}
\expandafter\def\csname m@scratch_procrustes@norm\endcsname{0.0}
\expandafter\def\csname m@scratch_procrustes@recall\endcsname{0.0}
\expandafter\def\csname m@scratch_procrustes@gauc\endcsname{+2.6}
\expandafter\def\csname m@scratch_procrustes@logloss\endcsname{-0.2}
\expandafter\def\csname m@scratch_procrustes@p2v_minutes\endcsname{0.0}
\expandafter\def\csname m@scratch_procrustes@dnn_minutes\endcsname{\NA}
\expandafter\def\csname m@scratch_unit@cosine\endcsname{0.0}
\expandafter\def\csname m@scratch_unit@l2\endcsname{-63.2}
\expandafter\def\csname m@scratch_unit@norm\endcsname{-62.5}
\expandafter\def\csname m@scratch_unit@recall\endcsname{0.0}
\expandafter\def\csname m@scratch_unit@gauc\endcsname{\NA}
\expandafter\def\csname m@scratch_unit@logloss\endcsname{\NA}
\expandafter\def\csname m@scratch_unit@p2v_minutes\endcsname{\NA}
\expandafter\def\csname m@scratch_unit@dnn_minutes\endcsname{\NA}
\expandafter\def\csname m@scratch_unit_op@cosine\endcsname{+93.4}
\expandafter\def\csname m@scratch_unit_op@l2\endcsname{-81.6}
\expandafter\def\csname m@scratch_unit_op@norm\endcsname{-62.5}
\expandafter\def\csname m@scratch_unit_op@recall\endcsname{0.0}
\expandafter\def\csname m@scratch_unit_op@gauc\endcsname{\NA}
\expandafter\def\csname m@scratch_unit_op@logloss\endcsname{\NA}
\expandafter\def\csname m@scratch_unit_op@p2v_minutes\endcsname{\NA}
\expandafter\def\csname m@scratch_unit_op@dnn_minutes\endcsname{\NA}
\expandafter\def\csname m@aligned_scratch_affine@cosine\endcsname{+124.9}
\expandafter\def\csname m@aligned_scratch_affine@l2\endcsname{-89.0}
\expandafter\def\csname m@aligned_scratch_affine@norm\endcsname{-26.7}
\expandafter\def\csname m@aligned_scratch_affine@recall\endcsname{-58.3}
\expandafter\def\csname m@aligned_scratch_affine@gauc\endcsname{+2.8}
\expandafter\def\csname m@aligned_scratch_affine@logloss\endcsname{-0.3}
\expandafter\def\csname m@aligned_scratch_affine@p2v_minutes\endcsname{\NA}
\expandafter\def\csname m@aligned_scratch_affine@dnn_minutes\endcsname{+87.1}
\expandafter\def\csname m@aligned_scratch_mlp@cosine\endcsname{+125.4}
\expandafter\def\csname m@aligned_scratch_mlp@l2\endcsname{-89.4}
\expandafter\def\csname m@aligned_scratch_mlp@norm\endcsname{-21.8}
\expandafter\def\csname m@aligned_scratch_mlp@recall\endcsname{-62.4}
\expandafter\def\csname m@aligned_scratch_mlp@gauc\endcsname{+2.5}
\expandafter\def\csname m@aligned_scratch_mlp@logloss\endcsname{-0.3}
\expandafter\def\csname m@aligned_scratch_mlp@p2v_minutes\endcsname{\NA}
\expandafter\def\csname m@aligned_scratch_mlp@dnn_minutes\endcsname{+221.2}
\expandafter\def\csname m@xsimgcl_procrustes@cosine\endcsname{-35.0}
\expandafter\def\csname m@xsimgcl_procrustes@l2\endcsname{+8.1}
\expandafter\def\csname m@xsimgcl_procrustes@norm\endcsname{-0.9}
\expandafter\def\csname m@xsimgcl_procrustes@recall\endcsname{+76.4}
\expandafter\def\csname m@xsimgcl_procrustes@gauc\endcsname{+2.2}
\expandafter\def\csname m@xsimgcl_procrustes@logloss\endcsname{-0.2}
\expandafter\def\csname m@xsimgcl_procrustes@p2v_minutes\endcsname{+22.2}
\expandafter\def\csname m@xsimgcl_procrustes@dnn_minutes\endcsname{\NA}
\expandafter\def\csname m@xsimgcl_unit@cosine\endcsname{-102.4}
\expandafter\def\csname m@xsimgcl_unit@l2\endcsname{-50.4}
\expandafter\def\csname m@xsimgcl_unit@norm\endcsname{-62.5}
\expandafter\def\csname m@xsimgcl_unit@recall\endcsname{+76.4}
\expandafter\def\csname m@xsimgcl_unit@gauc\endcsname{\NA}
\expandafter\def\csname m@xsimgcl_unit@logloss\endcsname{\NA}
\expandafter\def\csname m@xsimgcl_unit@p2v_minutes\endcsname{\NA}
\expandafter\def\csname m@xsimgcl_unit@dnn_minutes\endcsname{\NA}
\expandafter\def\csname m@xsimgcl_unit_op@cosine\endcsname{-35.0}
\expandafter\def\csname m@xsimgcl_unit_op@l2\endcsname{-58.7}
\expandafter\def\csname m@xsimgcl_unit_op@norm\endcsname{-62.5}
\expandafter\def\csname m@xsimgcl_unit_op@recall\endcsname{+76.4}
\expandafter\def\csname m@xsimgcl_unit_op@gauc\endcsname{\NA}
\expandafter\def\csname m@xsimgcl_unit_op@logloss\endcsname{\NA}
\expandafter\def\csname m@xsimgcl_unit_op@p2v_minutes\endcsname{\NA}
\expandafter\def\csname m@xsimgcl_unit_op@dnn_minutes\endcsname{\NA}
\expandafter\def\csname m@word2vec_idinit_raw@cosine\endcsname{+51.2}
\expandafter\def\csname m@word2vec_idinit_raw@l2\endcsname{-23.1}
\expandafter\def\csname m@word2vec_idinit_raw@norm\endcsname{0.0}
\expandafter\def\csname m@word2vec_idinit_raw@recall\endcsname{-0.4}
\expandafter\def\csname m@word2vec_idinit_raw@gauc\endcsname{+2.2}
\expandafter\def\csname m@word2vec_idinit_raw@logloss\endcsname{-0.6}
\expandafter\def\csname m@word2vec_idinit_raw@p2v_minutes\endcsname{+36.1}
\expandafter\def\csname m@word2vec_idinit_raw@dnn_minutes\endcsname{\NA}
\expandafter\def\csname m@word2vec_idinit_procrustes@cosine\endcsname{+98.4}
\expandafter\def\csname m@word2vec_idinit_procrustes@l2\endcsname{-52.1}
\expandafter\def\csname m@word2vec_idinit_procrustes@norm\endcsname{0.0}
\expandafter\def\csname m@word2vec_idinit_procrustes@recall\endcsname{-0.4}
\expandafter\def\csname m@word2vec_idinit_procrustes@gauc\endcsname{+2.7}
\expandafter\def\csname m@word2vec_idinit_procrustes@logloss\endcsname{-0.2}
\expandafter\def\csname m@word2vec_idinit_procrustes@p2v_minutes\endcsname{+36.1}
\expandafter\def\csname m@word2vec_idinit_procrustes@dnn_minutes\endcsname{\NA}
\expandafter\def\csname m@word2vec_idinit_unit@cosine\endcsname{+51.2}
\expandafter\def\csname m@word2vec_idinit_unit@l2\endcsname{-71.5}
\expandafter\def\csname m@word2vec_idinit_unit@norm\endcsname{-62.5}
\expandafter\def\csname m@word2vec_idinit_unit@recall\endcsname{-0.4}
\expandafter\def\csname m@word2vec_idinit_unit@gauc\endcsname{\NA}
\expandafter\def\csname m@word2vec_idinit_unit@logloss\endcsname{\NA}
\expandafter\def\csname m@word2vec_idinit_unit@p2v_minutes\endcsname{\NA}
\expandafter\def\csname m@word2vec_idinit_unit@dnn_minutes\endcsname{\NA}
\expandafter\def\csname m@word2vec_idinit_unit_op@cosine\endcsname{+98.4}
\expandafter\def\csname m@word2vec_idinit_unit_op@l2\endcsname{-83.0}
\expandafter\def\csname m@word2vec_idinit_unit_op@norm\endcsname{-62.5}
\expandafter\def\csname m@word2vec_idinit_unit_op@recall\endcsname{-0.4}
\expandafter\def\csname m@word2vec_idinit_unit_op@gauc\endcsname{+2.1}
\expandafter\def\csname m@word2vec_idinit_unit_op@logloss\endcsname{-0.4}
\expandafter\def\csname m@word2vec_idinit_unit_op@p2v_minutes\endcsname{\NA}
\expandafter\def\csname m@word2vec_idinit_unit_op@dnn_minutes\endcsname{+198.1}
\expandafter\def\csname m@aligned_warm@cosine\endcsname{+127.0}
\expandafter\def\csname m@aligned_warm@l2\endcsname{-83.7}
\expandafter\def\csname m@aligned_warm@norm\endcsname{+102.5}
\expandafter\def\csname m@aligned_warm@recall\endcsname{-2.3}
\expandafter\def\csname m@aligned_warm@gauc\endcsname{+3.5}
\expandafter\def\csname m@aligned_warm@logloss\endcsname{-0.4}
\expandafter\def\csname m@aligned_warm@p2v_minutes\endcsname{+59.4}
\expandafter\def\csname m@aligned_warm@dnn_minutes\endcsname{+54.2}
\expandafter\def\csname m@aligned_warm_norm@cosine\endcsname{+126.4}
\expandafter\def\csname m@aligned_warm_norm@l2\endcsname{-88.3}
\expandafter\def\csname m@aligned_warm_norm@norm\endcsname{+8.2}
\expandafter\def\csname m@aligned_warm_norm@recall\endcsname{+15.9}
\expandafter\def\csname m@aligned_warm_norm@gauc\endcsname{+4.0}
\expandafter\def\csname m@aligned_warm_norm@logloss\endcsname{-0.5}
\expandafter\def\csname m@aligned_warm_norm@p2v_minutes\endcsname{+49.4}
\expandafter\def\csname m@aligned_warm_norm@dnn_minutes\endcsname{+53.7}
\expandafter\def\csname m@aligned_warm_unit@cosine\endcsname{+127.0}
\expandafter\def\csname m@aligned_warm_unit@l2\endcsname{-97.2}
\expandafter\def\csname m@aligned_warm_unit@norm\endcsname{-62.5}
\expandafter\def\csname m@aligned_warm_unit@recall\endcsname{-2.3}
\expandafter\def\csname m@aligned_warm_unit@gauc\endcsname{+3.1}
\expandafter\def\csname m@aligned_warm_unit@logloss\endcsname{-0.5}
\expandafter\def\csname m@aligned_warm_unit@p2v_minutes\endcsname{+59.4}
\expandafter\def\csname m@aligned_warm_unit@dnn_minutes\endcsname{+62.8}
\expandafter\def\csname m@xsimgcl_idinit_raw@cosine\endcsname{+77.9}
\expandafter\def\csname m@xsimgcl_idinit_raw@l2\endcsname{-41.3}
\expandafter\def\csname m@xsimgcl_idinit_raw@norm\endcsname{-0.9}
\expandafter\def\csname m@xsimgcl_idinit_raw@recall\endcsname{+77.4}
\expandafter\def\csname m@xsimgcl_idinit_raw@gauc\endcsname{+1.9}
\expandafter\def\csname m@xsimgcl_idinit_raw@logloss\endcsname{-0.3}
\expandafter\def\csname m@xsimgcl_idinit_raw@p2v_minutes\endcsname{+15.5}
\expandafter\def\csname m@xsimgcl_idinit_raw@dnn_minutes\endcsname{\NA}
\expandafter\def\csname m@xsimgcl_idinit_procrustes@cosine\endcsname{+81.7}
\expandafter\def\csname m@xsimgcl_idinit_procrustes@l2\endcsname{-43.9}
\expandafter\def\csname m@xsimgcl_idinit_procrustes@norm\endcsname{-0.9}
\expandafter\def\csname m@xsimgcl_idinit_procrustes@recall\endcsname{+77.4}
\expandafter\def\csname m@xsimgcl_idinit_procrustes@gauc\endcsname{+2.4}
\expandafter\def\csname m@xsimgcl_idinit_procrustes@logloss\endcsname{-0.4}
\expandafter\def\csname m@xsimgcl_idinit_procrustes@p2v_minutes\endcsname{+15.5}
\expandafter\def\csname m@xsimgcl_idinit_procrustes@dnn_minutes\endcsname{\NA}
\expandafter\def\csname m@xsimgcl_idinit_unit@cosine\endcsname{+77.9}
\expandafter\def\csname m@xsimgcl_idinit_unit@l2\endcsname{-77.3}
\expandafter\def\csname m@xsimgcl_idinit_unit@norm\endcsname{-62.5}
\expandafter\def\csname m@xsimgcl_idinit_unit@recall\endcsname{+77.4}
\expandafter\def\csname m@xsimgcl_idinit_unit@gauc\endcsname{\NA}
\expandafter\def\csname m@xsimgcl_idinit_unit@logloss\endcsname{\NA}
\expandafter\def\csname m@xsimgcl_idinit_unit@p2v_minutes\endcsname{\NA}
\expandafter\def\csname m@xsimgcl_idinit_unit@dnn_minutes\endcsname{\NA}
\expandafter\def\csname m@xsimgcl_idinit_unit_op@cosine\endcsname{+81.7}
\expandafter\def\csname m@xsimgcl_idinit_unit_op@l2\endcsname{-78.2}
\expandafter\def\csname m@xsimgcl_idinit_unit_op@norm\endcsname{-62.5}
\expandafter\def\csname m@xsimgcl_idinit_unit_op@recall\endcsname{+77.4}
\expandafter\def\csname m@xsimgcl_idinit_unit_op@gauc\endcsname{+2.2}
\expandafter\def\csname m@xsimgcl_idinit_unit_op@logloss\endcsname{-0.3}
\expandafter\def\csname m@xsimgcl_idinit_unit_op@p2v_minutes\endcsname{\NA}
\expandafter\def\csname m@xsimgcl_idinit_unit_op@dnn_minutes\endcsname{+202.7}
\expandafter\def\csname m@sphere_raw@cosine\endcsname{-61.4}
\expandafter\def\csname m@sphere_raw@l2\endcsname{-55.0}
\expandafter\def\csname m@sphere_raw@norm\endcsname{-62.5}
\expandafter\def\csname m@sphere_raw@recall\endcsname{-6.3}
\expandafter\def\csname m@sphere_raw@gauc\endcsname{\NA}
\expandafter\def\csname m@sphere_raw@logloss\endcsname{\NA}
\expandafter\def\csname m@sphere_raw@p2v_minutes\endcsname{+64.5}
\expandafter\def\csname m@sphere_raw@dnn_minutes\endcsname{\NA}
\expandafter\def\csname m@sphere_procrustes@cosine\endcsname{-14.0}
\expandafter\def\csname m@sphere_procrustes@l2\endcsname{-61.6}
\expandafter\def\csname m@sphere_procrustes@norm\endcsname{-62.5}
\expandafter\def\csname m@sphere_procrustes@recall\endcsname{-6.3}
\expandafter\def\csname m@sphere_procrustes@gauc\endcsname{\NA}
\expandafter\def\csname m@sphere_procrustes@logloss\endcsname{\NA}
\expandafter\def\csname m@sphere_procrustes@p2v_minutes\endcsname{+64.5}
\expandafter\def\csname m@sphere_procrustes@dnn_minutes\endcsname{\NA}
\expandafter\def\csname m@sphere_idinit_raw@cosine\endcsname{+97.8}
\expandafter\def\csname m@sphere_idinit_raw@l2\endcsname{-82.4}
\expandafter\def\csname m@sphere_idinit_raw@norm\endcsname{-62.5}
\expandafter\def\csname m@sphere_idinit_raw@recall\endcsname{-3.9}
\expandafter\def\csname m@sphere_idinit_raw@gauc\endcsname{\NA}
\expandafter\def\csname m@sphere_idinit_raw@logloss\endcsname{\NA}
\expandafter\def\csname m@sphere_idinit_raw@p2v_minutes\endcsname{+67.4}
\expandafter\def\csname m@sphere_idinit_raw@dnn_minutes\endcsname{\NA}
\expandafter\def\csname m@sphere_idinit_procrustes@cosine\endcsname{+116.8}
\expandafter\def\csname m@sphere_idinit_procrustes@l2\endcsname{-89.4}
\expandafter\def\csname m@sphere_idinit_procrustes@norm\endcsname{-62.5}
\expandafter\def\csname m@sphere_idinit_procrustes@recall\endcsname{-3.9}
\expandafter\def\csname m@sphere_idinit_procrustes@gauc\endcsname{+1.8}
\expandafter\def\csname m@sphere_idinit_procrustes@logloss\endcsname{-0.2}
\expandafter\def\csname m@sphere_idinit_procrustes@p2v_minutes\endcsname{+67.4}
\expandafter\def\csname m@sphere_idinit_procrustes@dnn_minutes\endcsname{+117.9}
\expandafter\def\csname m@word2vec_idinit_temporal_raw@cosine\endcsname{+88.4}
\expandafter\def\csname m@word2vec_idinit_temporal_raw@l2\endcsname{-44.7}
\expandafter\def\csname m@word2vec_idinit_temporal_raw@norm\endcsname{\NA}
\expandafter\def\csname m@word2vec_idinit_temporal_raw@recall\endcsname{+3.8}
\expandafter\def\csname m@word2vec_idinit_temporal_raw@gauc\endcsname{\NA}
\expandafter\def\csname m@word2vec_idinit_temporal_raw@logloss\endcsname{\NA}
\expandafter\def\csname m@word2vec_idinit_temporal_raw@p2v_minutes\endcsname{+70.1}
\expandafter\def\csname m@word2vec_idinit_temporal_raw@dnn_minutes\endcsname{\NA}
\expandafter\def\csname m@word2vec_idinit_temporal_procrustes@cosine\endcsname{+110.9}
\expandafter\def\csname m@word2vec_idinit_temporal_procrustes@l2\endcsname{-63.5}
\expandafter\def\csname m@word2vec_idinit_temporal_procrustes@norm\endcsname{\NA}
\expandafter\def\csname m@word2vec_idinit_temporal_procrustes@recall\endcsname{+3.8}
\expandafter\def\csname m@word2vec_idinit_temporal_procrustes@gauc\endcsname{\NA}
\expandafter\def\csname m@word2vec_idinit_temporal_procrustes@logloss\endcsname{\NA}
\expandafter\def\csname m@word2vec_idinit_temporal_procrustes@p2v_minutes\endcsname{+70.1}
\expandafter\def\csname m@word2vec_idinit_temporal_procrustes@dnn_minutes\endcsname{\NA}
\expandafter\def\csname m@xsimgcl_idinit_temporal_raw@cosine\endcsname{+81.2}
\expandafter\def\csname m@xsimgcl_idinit_temporal_raw@l2\endcsname{-44.5}
\expandafter\def\csname m@xsimgcl_idinit_temporal_raw@norm\endcsname{\NA}
\expandafter\def\csname m@xsimgcl_idinit_temporal_raw@recall\endcsname{+77.3}
\expandafter\def\csname m@xsimgcl_idinit_temporal_raw@gauc\endcsname{\NA}
\expandafter\def\csname m@xsimgcl_idinit_temporal_raw@logloss\endcsname{\NA}
\expandafter\def\csname m@xsimgcl_idinit_temporal_raw@p2v_minutes\endcsname{+31.9}
\expandafter\def\csname m@xsimgcl_idinit_temporal_raw@dnn_minutes\endcsname{\NA}
\expandafter\def\csname m@xsimgcl_idinit_temporal_procrustes@cosine\endcsname{+84.5}
\expandafter\def\csname m@xsimgcl_idinit_temporal_procrustes@l2\endcsname{-46.7}
\expandafter\def\csname m@xsimgcl_idinit_temporal_procrustes@norm\endcsname{\NA}
\expandafter\def\csname m@xsimgcl_idinit_temporal_procrustes@recall\endcsname{+77.3}
\expandafter\def\csname m@xsimgcl_idinit_temporal_procrustes@gauc\endcsname{\NA}
\expandafter\def\csname m@xsimgcl_idinit_temporal_procrustes@logloss\endcsname{\NA}
\expandafter\def\csname m@xsimgcl_idinit_temporal_procrustes@p2v_minutes\endcsname{+31.9}
\expandafter\def\csname m@xsimgcl_idinit_temporal_procrustes@dnn_minutes\endcsname{\NA}
\expandafter\def\csname m@scratch_fixed_day7_raw@cosine\endcsname{0.0}
\expandafter\def\csname m@scratch_fixed_day7_raw@l2\endcsname{0.0}
\expandafter\def\csname m@scratch_fixed_day7_raw@norm\endcsname{\NA}
\expandafter\def\csname m@scratch_fixed_day7_raw@recall\endcsname{0.0}
\expandafter\def\csname m@scratch_fixed_day7_raw@gauc\endcsname{-0.2}
\expandafter\def\csname m@scratch_fixed_day7_raw@logloss\endcsname{+0.4}
\expandafter\def\csname m@scratch_fixed_day7_raw@p2v_minutes\endcsname{\NA}
\expandafter\def\csname m@scratch_fixed_day7_raw@dnn_minutes\endcsname{-29.6}
\expandafter\def\csname m@scratch_fixed_day7_procrustes@cosine\endcsname{+93.4}
\expandafter\def\csname m@scratch_fixed_day7_procrustes@l2\endcsname{-48.2}
\expandafter\def\csname m@scratch_fixed_day7_procrustes@norm\endcsname{\NA}
\expandafter\def\csname m@scratch_fixed_day7_procrustes@recall\endcsname{0.0}
\expandafter\def\csname m@scratch_fixed_day7_procrustes@gauc\endcsname{+3.9}
\expandafter\def\csname m@scratch_fixed_day7_procrustes@logloss\endcsname{-0.3}
\expandafter\def\csname m@scratch_fixed_day7_procrustes@p2v_minutes\endcsname{\NA}
\expandafter\def\csname m@scratch_fixed_day7_procrustes@dnn_minutes\endcsname{+69.7}
\expandafter\def\csname m@xsimgcl_fixed_day7_procrustes@cosine\endcsname{-35.0}
\expandafter\def\csname m@xsimgcl_fixed_day7_procrustes@l2\endcsname{+8.1}
\expandafter\def\csname m@xsimgcl_fixed_day7_procrustes@norm\endcsname{\NA}
\expandafter\def\csname m@xsimgcl_fixed_day7_procrustes@recall\endcsname{+76.4}
\expandafter\def\csname m@xsimgcl_fixed_day7_procrustes@gauc\endcsname{+1.2}
\expandafter\def\csname m@xsimgcl_fixed_day7_procrustes@logloss\endcsname{-0.2}
\expandafter\def\csname m@xsimgcl_fixed_day7_procrustes@p2v_minutes\endcsname{\NA}
\expandafter\def\csname m@xsimgcl_fixed_day7_procrustes@dnn_minutes\endcsname{+71.8}
\expandafter\def\csname m@noise025@cosine\endcsname{+93.4}
\expandafter\def\csname m@noise025@l2\endcsname{-48.2}
\expandafter\def\csname m@noise025@norm\endcsname{0.0}
\expandafter\def\csname m@noise025@recall\endcsname{0.0}
\expandafter\def\csname m@noise025@gauc\endcsname{+3.0}
\expandafter\def\csname m@noise025@logloss\endcsname{-0.2}
\expandafter\def\csname m@noise025@p2v_minutes\endcsname{0.0}
\expandafter\def\csname m@noise025@dnn_minutes\endcsname{+121.1}
\expandafter\def\csname m@noise050@cosine\endcsname{+93.4}
\expandafter\def\csname m@noise050@l2\endcsname{-48.2}
\expandafter\def\csname m@noise050@norm\endcsname{0.0}
\expandafter\def\csname m@noise050@recall\endcsname{0.0}
\expandafter\def\csname m@noise050@gauc\endcsname{+3.3}
\expandafter\def\csname m@noise050@logloss\endcsname{-0.3}
\expandafter\def\csname m@noise050@p2v_minutes\endcsname{0.0}
\expandafter\def\csname m@noise050@dnn_minutes\endcsname{+101.6}
\expandafter\def\csname m@noise100@cosine\endcsname{+93.4}
\expandafter\def\csname m@noise100@l2\endcsname{-48.2}
\expandafter\def\csname m@noise100@norm\endcsname{0.0}
\expandafter\def\csname m@noise100@recall\endcsname{0.0}
\expandafter\def\csname m@noise100@gauc\endcsname{+2.9}
\expandafter\def\csname m@noise100@logloss\endcsname{-0.3}
\expandafter\def\csname m@noise100@p2v_minutes\endcsname{0.0}
\expandafter\def\csname m@noise100@dnn_minutes\endcsname{+93.9}
\expandafter\def\csname m@history050@cosine\endcsname{+93.4}
\expandafter\def\csname m@history050@l2\endcsname{-48.2}
\expandafter\def\csname m@history050@norm\endcsname{0.0}
\expandafter\def\csname m@history050@recall\endcsname{0.0}
\expandafter\def\csname m@history050@gauc\endcsname{+2.6}
\expandafter\def\csname m@history050@logloss\endcsname{-0.2}
\expandafter\def\csname m@history050@p2v_minutes\endcsname{0.0}
\expandafter\def\csname m@history050@dnn_minutes\endcsname{+727.0}
\expandafter\def\csname m@history_all@cosine\endcsname{+93.4}
\expandafter\def\csname m@history_all@l2\endcsname{-48.2}
\expandafter\def\csname m@history_all@norm\endcsname{0.0}
\expandafter\def\csname m@history_all@recall\endcsname{0.0}
\expandafter\def\csname m@history_all@gauc\endcsname{+2.0}
\expandafter\def\csname m@history_all@logloss\endcsname{-0.3}
\expandafter\def\csname m@history_all@p2v_minutes\endcsname{0.0}
\expandafter\def\csname m@history_all@dnn_minutes\endcsname{+776.4}
\expandafter\def\csname m@aligned_warm_norm_noise050@cosine\endcsname{+126.4}
\expandafter\def\csname m@aligned_warm_norm_noise050@l2\endcsname{-88.3}
\expandafter\def\csname m@aligned_warm_norm_noise050@norm\endcsname{+8.2}
\expandafter\def\csname m@aligned_warm_norm_noise050@recall\endcsname{+15.9}
\expandafter\def\csname m@aligned_warm_norm_noise050@gauc\endcsname{+3.8}
\expandafter\def\csname m@aligned_warm_norm_noise050@logloss\endcsname{-0.5}
\expandafter\def\csname m@aligned_warm_norm_noise050@p2v_minutes\endcsname{+49.4}
\expandafter\def\csname m@aligned_warm_norm_noise050@dnn_minutes\endcsname{+12.5}
\expandafter\def\csname m@history_all_raw@cosine\endcsname{0.0}
\expandafter\def\csname m@history_all_raw@l2\endcsname{0.0}
\expandafter\def\csname m@history_all_raw@norm\endcsname{0.0}
\expandafter\def\csname m@history_all_raw@recall\endcsname{0.0}
\expandafter\def\csname m@history_all_raw@gauc\endcsname{0.0}
\expandafter\def\csname m@history_all_raw@logloss\endcsname{+0.1}
\expandafter\def\csname m@history_all_raw@p2v_minutes\endcsname{0.0}
\expandafter\def\csname m@history_all_raw@dnn_minutes\endcsname{+517.3}
\expandafter\def\csname m@word2vec_idinit_procrustes_noise050@cosine\endcsname{+98.4}
\expandafter\def\csname m@word2vec_idinit_procrustes_noise050@l2\endcsname{-52.1}
\expandafter\def\csname m@word2vec_idinit_procrustes_noise050@norm\endcsname{0.0}
\expandafter\def\csname m@word2vec_idinit_procrustes_noise050@recall\endcsname{-0.4}
\expandafter\def\csname m@word2vec_idinit_procrustes_noise050@gauc\endcsname{+2.8}
\expandafter\def\csname m@word2vec_idinit_procrustes_noise050@logloss\endcsname{-0.3}
\expandafter\def\csname m@word2vec_idinit_procrustes_noise050@p2v_minutes\endcsname{+36.1}
\expandafter\def\csname m@word2vec_idinit_procrustes_noise050@dnn_minutes\endcsname{+95.2}
\expandafter\def\csname m@xsimgcl_idinit_procrustes_noise050@cosine\endcsname{+81.7}
\expandafter\def\csname m@xsimgcl_idinit_procrustes_noise050@l2\endcsname{-43.9}
\expandafter\def\csname m@xsimgcl_idinit_procrustes_noise050@norm\endcsname{-0.9}
\expandafter\def\csname m@xsimgcl_idinit_procrustes_noise050@recall\endcsname{+77.4}
\expandafter\def\csname m@xsimgcl_idinit_procrustes_noise050@gauc\endcsname{+2.2}
\expandafter\def\csname m@xsimgcl_idinit_procrustes_noise050@logloss\endcsname{-0.4}
\expandafter\def\csname m@xsimgcl_idinit_procrustes_noise050@p2v_minutes\endcsname{+15.5}
\expandafter\def\csname m@xsimgcl_idinit_procrustes_noise050@dnn_minutes\endcsname{+104.3}
\expandafter\def\csname m@word2vec_idinit_unit_op_noise050@cosine\endcsname{+98.4}
\expandafter\def\csname m@word2vec_idinit_unit_op_noise050@l2\endcsname{-83.0}
\expandafter\def\csname m@word2vec_idinit_unit_op_noise050@norm\endcsname{-62.5}
\expandafter\def\csname m@word2vec_idinit_unit_op_noise050@recall\endcsname{-0.4}
\expandafter\def\csname m@word2vec_idinit_unit_op_noise050@gauc\endcsname{+1.5}
\expandafter\def\csname m@word2vec_idinit_unit_op_noise050@logloss\endcsname{-0.4}
\expandafter\def\csname m@word2vec_idinit_unit_op_noise050@p2v_minutes\endcsname{\NA}
\expandafter\def\csname m@word2vec_idinit_unit_op_noise050@dnn_minutes\endcsname{+209.6}
\expandafter\def\csname m@xsimgcl_idinit_unit_op_noise050@cosine\endcsname{+81.7}
\expandafter\def\csname m@xsimgcl_idinit_unit_op_noise050@l2\endcsname{-78.2}
\expandafter\def\csname m@xsimgcl_idinit_unit_op_noise050@norm\endcsname{-62.5}
\expandafter\def\csname m@xsimgcl_idinit_unit_op_noise050@recall\endcsname{+77.4}
\expandafter\def\csname m@xsimgcl_idinit_unit_op_noise050@gauc\endcsname{+2.0}
\expandafter\def\csname m@xsimgcl_idinit_unit_op_noise050@logloss\endcsname{-0.3}
\expandafter\def\csname m@xsimgcl_idinit_unit_op_noise050@p2v_minutes\endcsname{\NA}
\expandafter\def\csname m@xsimgcl_idinit_unit_op_noise050@dnn_minutes\endcsname{+203.0}
\expandafter\def\csname m@sphere_idinit_procrustes_noise050@cosine\endcsname{+116.8}
\expandafter\def\csname m@sphere_idinit_procrustes_noise050@l2\endcsname{-89.4}
\expandafter\def\csname m@sphere_idinit_procrustes_noise050@norm\endcsname{-62.5}
\expandafter\def\csname m@sphere_idinit_procrustes_noise050@recall\endcsname{-3.9}
\expandafter\def\csname m@sphere_idinit_procrustes_noise050@gauc\endcsname{+2.3}
\expandafter\def\csname m@sphere_idinit_procrustes_noise050@logloss\endcsname{-0.3}
\expandafter\def\csname m@sphere_idinit_procrustes_noise050@p2v_minutes\endcsname{+67.4}
\expandafter\def\csname m@sphere_idinit_procrustes_noise050@dnn_minutes\endcsname{+115.2}
\expandafter\def\csname m@word2vec_single@cosine\endcsname{\NA}
\expandafter\def\csname m@word2vec_single@l2\endcsname{\NA}
\expandafter\def\csname m@word2vec_single@norm\endcsname{\NA}
\expandafter\def\csname m@word2vec_single@recall\endcsname{\NA}
\expandafter\def\csname m@word2vec_single@gauc\endcsname{\NA}
\expandafter\def\csname m@word2vec_single@logloss\endcsname{\NA}
\expandafter\def\csname m@word2vec_single@p2v_minutes\endcsname{+700.3}
\expandafter\def\csname m@word2vec_single@dnn_minutes\endcsname{\NA}
\expandafter\def\csname m@sync_1@cosine\endcsname{\NA}
\expandafter\def\csname m@sync_1@l2\endcsname{\NA}
\expandafter\def\csname m@sync_1@norm\endcsname{\NA}
\expandafter\def\csname m@sync_1@recall\endcsname{-84.9}
\expandafter\def\csname m@sync_1@gauc\endcsname{\NA}
\expandafter\def\csname m@sync_1@logloss\endcsname{\NA}
\expandafter\def\csname m@sync_1@p2v_minutes\endcsname{\NA}
\expandafter\def\csname m@sync_1@dnn_minutes\endcsname{\NA}
\expandafter\def\csname m@sync_2@cosine\endcsname{\NA}
\expandafter\def\csname m@sync_2@l2\endcsname{\NA}
\expandafter\def\csname m@sync_2@norm\endcsname{\NA}
\expandafter\def\csname m@sync_2@recall\endcsname{-58.4}
\expandafter\def\csname m@sync_2@gauc\endcsname{\NA}
\expandafter\def\csname m@sync_2@logloss\endcsname{\NA}
\expandafter\def\csname m@sync_2@p2v_minutes\endcsname{\NA}
\expandafter\def\csname m@sync_2@dnn_minutes\endcsname{\NA}
\expandafter\def\csname m@warm_sphere@cosine\endcsname{+126.9}
\expandafter\def\csname m@warm_sphere@l2\endcsname{-96.9}
\expandafter\def\csname m@warm_sphere@norm\endcsname{-62.5}
\expandafter\def\csname m@warm_sphere@recall\endcsname{+47.1}
\expandafter\def\csname m@warm_sphere@gauc\endcsname{+3.3}
\expandafter\def\csname m@warm_sphere@logloss\endcsname{-0.3}
\expandafter\def\csname m@warm_sphere@p2v_minutes\endcsname{-5.1}
\expandafter\def\csname m@warm_sphere@dnn_minutes\endcsname{-27.9}
\expandafter\def\csname e@word2vec@adjacent@input@cosine\endcsname{0.0}
\expandafter\def\csname e@word2vec@adjacent@input@l2\endcsname{0.0}
\expandafter\def\csname e@word2vec@adjacent@random@cosine\endcsname{-7.6}
\expandafter\def\csname e@word2vec@adjacent@random@l2\endcsname{-43.9}
\expandafter\def\csname e@word2vec@adjacent@svd@cosine\endcsname{+29.6}
\expandafter\def\csname e@word2vec@adjacent@svd@l2\endcsname{-24.8}
\expandafter\def\csname e@word2vec@adjacent@untrained@cosine\endcsname{+61.0}
\expandafter\def\csname e@word2vec@adjacent@untrained@l2\endcsname{-39.0}
\expandafter\def\csname e@word2vec@adjacent@encoder@cosine\endcsname{+80.3}
\expandafter\def\csname e@word2vec@adjacent@encoder@l2\endcsname{-63.4}
\expandafter\def\csname e@word2vec@validation@input@cosine\endcsname{0.0}
\expandafter\def\csname e@word2vec@validation@input@l2\endcsname{0.0}
\expandafter\def\csname e@word2vec@validation@random@cosine\endcsname{-10.5}
\expandafter\def\csname e@word2vec@validation@random@l2\endcsname{-44.3}
\expandafter\def\csname e@word2vec@validation@svd@cosine\endcsname{+35.0}
\expandafter\def\csname e@word2vec@validation@svd@l2\endcsname{-27.8}
\expandafter\def\csname e@word2vec@validation@untrained@cosine\endcsname{+67.5}
\expandafter\def\csname e@word2vec@validation@untrained@l2\endcsname{-39.7}
\expandafter\def\csname e@word2vec@validation@encoder@cosine\endcsname{+68.9}
\expandafter\def\csname e@word2vec@validation@encoder@l2\endcsname{-59.0}
\expandafter\def\csname e@word2vec@test@input@cosine\endcsname{0.0}
\expandafter\def\csname e@word2vec@test@input@l2\endcsname{0.0}
\expandafter\def\csname e@word2vec@test@random@cosine\endcsname{-12.9}
\expandafter\def\csname e@word2vec@test@random@l2\endcsname{-43.3}
\expandafter\def\csname e@word2vec@test@svd@cosine\endcsname{+40.1}
\expandafter\def\csname e@word2vec@test@svd@l2\endcsname{-29.3}
\expandafter\def\csname e@word2vec@test@untrained@cosine\endcsname{+72.4}
\expandafter\def\csname e@word2vec@test@untrained@l2\endcsname{-40.8}
\expandafter\def\csname e@word2vec@test@encoder@cosine\endcsname{+72.6}
\expandafter\def\csname e@word2vec@test@encoder@l2\endcsname{-57.5}
\expandafter\def\csname e@word2vec_op@adjacent@input@cosine\endcsname{+96.2}
\expandafter\def\csname e@word2vec_op@adjacent@input@l2\endcsname{-49.2}
\expandafter\def\csname e@word2vec_op@adjacent@random@cosine\endcsname{+94.3}
\expandafter\def\csname e@word2vec_op@adjacent@random@l2\endcsname{-71.3}
\expandafter\def\csname e@word2vec_op@adjacent@svd@cosine\endcsname{+113.2}
\expandafter\def\csname e@word2vec_op@adjacent@svd@l2\endcsname{-67.7}
\expandafter\def\csname e@word2vec_op@adjacent@untrained@cosine\endcsname{+108.2}
\expandafter\def\csname e@word2vec_op@adjacent@untrained@l2\endcsname{-67.2}
\expandafter\def\csname e@word2vec_op@adjacent@encoder@cosine\endcsname{+117.7}
\expandafter\def\csname e@word2vec_op@adjacent@encoder@l2\endcsname{-79.3}
\expandafter\def\csname e@word2vec_op@validation@input@cosine\endcsname{+95.3}
\expandafter\def\csname e@word2vec_op@validation@input@l2\endcsname{-45.0}
\expandafter\def\csname e@word2vec_op@validation@random@cosine\endcsname{+92.5}
\expandafter\def\csname e@word2vec_op@validation@random@l2\endcsname{-68.9}
\expandafter\def\csname e@word2vec_op@validation@svd@cosine\endcsname{+116.1}
\expandafter\def\csname e@word2vec_op@validation@svd@l2\endcsname{-65.2}
\expandafter\def\csname e@word2vec_op@validation@untrained@cosine\endcsname{+110.0}
\expandafter\def\csname e@word2vec_op@validation@untrained@l2\endcsname{-64.8}
\expandafter\def\csname e@word2vec_op@validation@encoder@cosine\endcsname{+120.4}
\expandafter\def\csname e@word2vec_op@validation@encoder@l2\endcsname{-77.2}
\expandafter\def\csname e@word2vec_op@test@input@cosine\endcsname{+104.4}
\expandafter\def\csname e@word2vec_op@test@input@l2\endcsname{-44.6}
\expandafter\def\csname e@word2vec_op@test@random@cosine\endcsname{+101.5}
\expandafter\def\csname e@word2vec_op@test@random@l2\endcsname{-68.7}
\expandafter\def\csname e@word2vec_op@test@svd@cosine\endcsname{+127.4}
\expandafter\def\csname e@word2vec_op@test@svd@l2\endcsname{-65.1}
\expandafter\def\csname e@word2vec_op@test@untrained@cosine\endcsname{+120.6}
\expandafter\def\csname e@word2vec_op@test@untrained@l2\endcsname{-64.7}
\expandafter\def\csname e@word2vec_op@test@encoder@cosine\endcsname{+131.4}
\expandafter\def\csname e@word2vec_op@test@encoder@l2\endcsname{-76.5}
\expandafter\def\csname e@warm_norm@adjacent@input@cosine\endcsname{+127.6}
\expandafter\def\csname e@warm_norm@adjacent@input@l2\endcsname{-86.4}
\expandafter\def\csname e@warm_norm@adjacent@random@cosine\endcsname{+127.5}
\expandafter\def\csname e@warm_norm@adjacent@random@l2\endcsname{-92.4}
\expandafter\def\csname e@warm_norm@adjacent@svd@cosine\endcsname{+128.3}
\expandafter\def\csname e@warm_norm@adjacent@svd@l2\endcsname{-90.5}
\expandafter\def\csname e@warm_norm@adjacent@untrained@cosine\endcsname{+127.8}
\expandafter\def\csname e@warm_norm@adjacent@untrained@l2\endcsname{-91.2}
\expandafter\def\csname e@warm_norm@adjacent@encoder@cosine\endcsname{+127.7}
\expandafter\def\csname e@warm_norm@adjacent@encoder@l2\endcsname{-91.9}
\expandafter\def\csname e@warm_norm@validation@input@cosine\endcsname{+125.6}
\expandafter\def\csname e@warm_norm@validation@input@l2\endcsname{-72.0}
\expandafter\def\csname e@warm_norm@validation@random@cosine\endcsname{+124.9}
\expandafter\def\csname e@warm_norm@validation@random@l2\endcsname{-84.2}
\expandafter\def\csname e@warm_norm@validation@svd@cosine\endcsname{+129.5}
\expandafter\def\csname e@warm_norm@validation@svd@l2\endcsname{-80.9}
\expandafter\def\csname e@warm_norm@validation@untrained@cosine\endcsname{+127.1}
\expandafter\def\csname e@warm_norm@validation@untrained@l2\endcsname{-82.1}
\expandafter\def\csname e@warm_norm@validation@encoder@cosine\endcsname{+127.0}
\expandafter\def\csname e@warm_norm@validation@encoder@l2\endcsname{-83.9}
\expandafter\def\csname e@warm_norm@test@input@cosine\endcsname{+135.3}
\expandafter\def\csname e@warm_norm@test@input@l2\endcsname{-67.4}
\expandafter\def\csname e@warm_norm@test@random@cosine\endcsname{+134.4}
\expandafter\def\csname e@warm_norm@test@random@l2\endcsname{-81.7}
\expandafter\def\csname e@warm_norm@test@svd@cosine\endcsname{+140.7}
\expandafter\def\csname e@warm_norm@test@svd@l2\endcsname{-77.8}
\expandafter\def\csname e@warm_norm@test@untrained@cosine\endcsname{+137.4}
\expandafter\def\csname e@warm_norm@test@untrained@l2\endcsname{-79.4}
\expandafter\def\csname e@warm_norm@test@encoder@cosine\endcsname{+137.1}
\expandafter\def\csname e@warm_norm@test@encoder@l2\endcsname{-81.4}
\expandafter\def\csname e@word2vec_idinit_op@adjacent@input@cosine\endcsname{+101.3}
\expandafter\def\csname e@word2vec_idinit_op@adjacent@input@l2\endcsname{-53.2}
\expandafter\def\csname e@word2vec_idinit_op@adjacent@random@cosine\endcsname{+103.6}
\expandafter\def\csname e@word2vec_idinit_op@adjacent@random@l2\endcsname{-73.5}
\expandafter\def\csname e@word2vec_idinit_op@adjacent@svd@cosine\endcsname{+114.8}
\expandafter\def\csname e@word2vec_idinit_op@adjacent@svd@l2\endcsname{-69.4}
\expandafter\def\csname e@word2vec_idinit_op@adjacent@untrained@cosine\endcsname{+113.3}
\expandafter\def\csname e@word2vec_idinit_op@adjacent@untrained@l2\endcsname{-70.1}
\expandafter\def\csname e@word2vec_idinit_op@adjacent@encoder@cosine\endcsname{+117.7}
\expandafter\def\csname e@word2vec_idinit_op@adjacent@encoder@l2\endcsname{-79.6}
\expandafter\def\csname e@word2vec_idinit_op@validation@input@cosine\endcsname{+100.3}
\expandafter\def\csname e@word2vec_idinit_op@validation@input@l2\endcsname{-48.6}
\expandafter\def\csname e@word2vec_idinit_op@validation@random@cosine\endcsname{+103.1}
\expandafter\def\csname e@word2vec_idinit_op@validation@random@l2\endcsname{-70.8}
\expandafter\def\csname e@word2vec_idinit_op@validation@svd@cosine\endcsname{+117.6}
\expandafter\def\csname e@word2vec_idinit_op@validation@svd@l2\endcsname{-66.7}
\expandafter\def\csname e@word2vec_idinit_op@validation@untrained@cosine\endcsname{+115.4}
\expandafter\def\csname e@word2vec_idinit_op@validation@untrained@l2\endcsname{-67.4}
\expandafter\def\csname e@word2vec_idinit_op@validation@encoder@cosine\endcsname{+119.8}
\expandafter\def\csname e@word2vec_idinit_op@validation@encoder@l2\endcsname{-77.0}
\expandafter\def\csname e@word2vec_idinit_op@test@input@cosine\endcsname{+109.3}
\expandafter\def\csname e@word2vec_idinit_op@test@input@l2\endcsname{-47.8}
\expandafter\def\csname e@word2vec_idinit_op@test@random@cosine\endcsname{+112.5}
\expandafter\def\csname e@word2vec_idinit_op@test@random@l2\endcsname{-70.6}
\expandafter\def\csname e@word2vec_idinit_op@test@svd@cosine\endcsname{+128.9}
\expandafter\def\csname e@word2vec_idinit_op@test@svd@l2\endcsname{-66.4}
\expandafter\def\csname e@word2vec_idinit_op@test@untrained@cosine\endcsname{+126.0}
\expandafter\def\csname e@word2vec_idinit_op@test@untrained@l2\endcsname{-66.8}
\expandafter\def\csname e@word2vec_idinit_op@test@encoder@cosine\endcsname{+130.9}
\expandafter\def\csname e@word2vec_idinit_op@test@encoder@l2\endcsname{-76.6}
\expandafter\def\csname e@xsimgcl_idinit_op@adjacent@input@cosine\endcsname{+85.0}
\expandafter\def\csname e@xsimgcl_idinit_op@adjacent@input@l2\endcsname{-43.8}
\expandafter\def\csname e@xsimgcl_idinit_op@adjacent@random@cosine\endcsname{+84.5}
\expandafter\def\csname e@xsimgcl_idinit_op@adjacent@random@l2\endcsname{-68.4}
\expandafter\def\csname e@xsimgcl_idinit_op@adjacent@svd@cosine\endcsname{+96.3}
\expandafter\def\csname e@xsimgcl_idinit_op@adjacent@svd@l2\endcsname{-68.9}
\expandafter\def\csname e@xsimgcl_idinit_op@adjacent@untrained@cosine\endcsname{+100.2}
\expandafter\def\csname e@xsimgcl_idinit_op@adjacent@untrained@l2\endcsname{-60.9}
\expandafter\def\csname e@xsimgcl_idinit_op@adjacent@encoder@cosine\endcsname{+108.7}
\expandafter\def\csname e@xsimgcl_idinit_op@adjacent@encoder@l2\endcsname{-76.0}
\expandafter\def\csname e@xsimgcl_idinit_op@validation@input@cosine\endcsname{+80.3}
\expandafter\def\csname e@xsimgcl_idinit_op@validation@input@l2\endcsname{-38.2}
\expandafter\def\csname e@xsimgcl_idinit_op@validation@random@cosine\endcsname{+79.7}
\expandafter\def\csname e@xsimgcl_idinit_op@validation@random@l2\endcsname{-65.3}
\expandafter\def\csname e@xsimgcl_idinit_op@validation@svd@cosine\endcsname{+93.9}
\expandafter\def\csname e@xsimgcl_idinit_op@validation@svd@l2\endcsname{-65.6}
\expandafter\def\csname e@xsimgcl_idinit_op@validation@untrained@cosine\endcsname{+99.9}
\expandafter\def\csname e@xsimgcl_idinit_op@validation@untrained@l2\endcsname{-57.7}
\expandafter\def\csname e@xsimgcl_idinit_op@validation@encoder@cosine\endcsname{+109.3}
\expandafter\def\csname e@xsimgcl_idinit_op@validation@encoder@l2\endcsname{-73.6}
\expandafter\def\csname e@xsimgcl_idinit_op@test@input@cosine\endcsname{+83.6}
\expandafter\def\csname e@xsimgcl_idinit_op@test@input@l2\endcsname{-35.8}
\expandafter\def\csname e@xsimgcl_idinit_op@test@random@cosine\endcsname{+83.1}
\expandafter\def\csname e@xsimgcl_idinit_op@test@random@l2\endcsname{-64.0}
\expandafter\def\csname e@xsimgcl_idinit_op@test@svd@cosine\endcsname{+99.3}
\expandafter\def\csname e@xsimgcl_idinit_op@test@svd@l2\endcsname{-64.2}
\expandafter\def\csname e@xsimgcl_idinit_op@test@untrained@cosine\endcsname{+108.0}
\expandafter\def\csname e@xsimgcl_idinit_op@test@untrained@l2\endcsname{-56.9}
\expandafter\def\csname e@xsimgcl_idinit_op@test@encoder@cosine\endcsname{+118.2}
\expandafter\def\csname e@xsimgcl_idinit_op@test@encoder@l2\endcsname{-72.8}
\expandafter\def\csname e@sphere_idinit_op@adjacent@input@cosine\endcsname{+119.1}
\expandafter\def\csname e@sphere_idinit_op@adjacent@input@l2\endcsname{-89.1}
\expandafter\def\csname e@sphere_idinit_op@adjacent@random@cosine\endcsname{+119.1}
\expandafter\def\csname e@sphere_idinit_op@adjacent@random@l2\endcsname{-93.8}
\expandafter\def\csname e@sphere_idinit_op@adjacent@svd@cosine\endcsname{+123.3}
\expandafter\def\csname e@sphere_idinit_op@adjacent@svd@l2\endcsname{-93.9}
\expandafter\def\csname e@sphere_idinit_op@adjacent@untrained@cosine\endcsname{+124.9}
\expandafter\def\csname e@sphere_idinit_op@adjacent@untrained@l2\endcsname{-84.9}
\expandafter\def\csname e@sphere_idinit_op@adjacent@encoder@cosine\endcsname{+123.4}
\expandafter\def\csname e@sphere_idinit_op@adjacent@encoder@l2\endcsname{-85.0}
\expandafter\def\csname e@sphere_idinit_op@validation@input@cosine\endcsname{+120.1}
\expandafter\def\csname e@sphere_idinit_op@validation@input@l2\endcsname{-87.1}
\expandafter\def\csname e@sphere_idinit_op@validation@random@cosine\endcsname{+120.1}
\expandafter\def\csname e@sphere_idinit_op@validation@random@l2\endcsname{-92.8}
\expandafter\def\csname e@sphere_idinit_op@validation@svd@cosine\endcsname{+125.8}
\expandafter\def\csname e@sphere_idinit_op@validation@svd@l2\endcsname{-92.7}
\expandafter\def\csname e@sphere_idinit_op@validation@untrained@cosine\endcsname{+128.7}
\expandafter\def\csname e@sphere_idinit_op@validation@untrained@l2\endcsname{-82.4}
\expandafter\def\csname e@sphere_idinit_op@validation@encoder@cosine\endcsname{+125.6}
\expandafter\def\csname e@sphere_idinit_op@validation@encoder@l2\endcsname{-81.6}
\expandafter\def\csname e@sphere_idinit_op@test@input@cosine\endcsname{+130.2}
\expandafter\def\csname e@sphere_idinit_op@test@input@l2\endcsname{-86.5}
\expandafter\def\csname e@sphere_idinit_op@test@random@cosine\endcsname{+130.0}
\expandafter\def\csname e@sphere_idinit_op@test@random@l2\endcsname{-92.4}
\expandafter\def\csname e@sphere_idinit_op@test@svd@cosine\endcsname{+137.2}
\expandafter\def\csname e@sphere_idinit_op@test@svd@l2\endcsname{-92.4}
\expandafter\def\csname e@sphere_idinit_op@test@untrained@cosine\endcsname{+140.6}
\expandafter\def\csname e@sphere_idinit_op@test@untrained@l2\endcsname{-81.7}
\expandafter\def\csname e@sphere_idinit_op@test@encoder@cosine\endcsname{+137.2}
\expandafter\def\csname e@sphere_idinit_op@test@encoder@l2\endcsname{-81.2}
\title{\vspace{-12mm}\textbf{Cross-Day Compatibility of Daily Item Embeddings}\\[5pt]\large An Empirical Study of Alignment, Scale Management,\\and CTR Adaptation}
\author{Yicheng Wang\\\normalsize LY Corporation}
\date{October 9, 2026}
\begin{document}
\maketitle
\begin{abstract}
Daily item co-occurrence embeddings are paired with next-day behavior without future-behavior leakage. Multi-day click-through rate (CTR) training therefore mixes versions whose directions and norms may differ for the same item. We compare Word2Vec and an item-graph XSimGCL adaptation across initialization, inheritance, post-processing, training constraints, and downstream augmentation. The inventory contains 48 configurations, including 32 common-protocol CTR evaluations. Observed metrics are reported as relative percentage changes from the corresponding Word2Vec baseline (0\%), with raw values and exact dataset sizes withheld. Three findings emerge. First, multiple upstream interventions improve compatibility and CTR. ID-init and orthogonal alignment increase cosine by 51.2\% and 93.4\%, and GAUC by 2.2\% and 2.6\%, relative to ordinary Word2Vec. Retrieval and geometry remain insufficient downstream surrogates. Second, warm starts perform strongly but accumulate scale. The day-matched norm increase over Word2Vec reaches 102.5\% after eight days; daily global rescaling limits it to 8.2\% and attains the highest observed GAUC increase, 4.0\%. Third, CTR learning through a dimensionality-reducing encoder further adapts to residual drift. Against dimension-matched random-projection, SVD, and untrained-network controls, trained encoders attain the highest cross-day cosine for four of six upstream methods. The other two rank third and second in train-to-test cosine. These industrial results support complementary roles for upstream stabilization, inherited-scale management, and downstream task learning.
\end{abstract}
\noindent\textbf{Keywords:} item co-occurrence embeddings; cross-day compatibility; warm starts; scale management; click-through rate prediction.

\section{Introduction}
Item co-occurrence embeddings capture relationships in behavioral sequences and provide compact features for recommendation and advertising. Word2Vec with skip-gram negative sampling (SGNS), and its item-based applications, learn these relationships from observed context pairs~\cite{mikolov2013distributed,barkan2016item2vec}. Graph alternatives aggregate neighborhood information and use contrastive objectives~\cite{he2020lightgcn,yu2022xsimgcl}. Both approaches require updates as behavior and the item catalog change.

Our pipeline pairs an embedding trained on information available through day $t$ with exposure and click logs from day $t+1$. This pairing incorporates recent information while excluding future behavioral observations from each example's embedding features. A CTR model trained on several days of logs consequently receives multiple daily embedding versions. The same item can have different directions and norms across those versions, so good retrieval within each version does not ensure that the versions function as compatible features for a shared downstream network.

Cross-day differences can reflect changing behavior, initialization, optimization, and non-identifiability of coordinates. For example, applying the same orthogonal transformation to both SGNS tables preserves their inner products. Independently trained versions can thus represent useful within-version relationships in different coordinate systems, even when their rolling data windows overlap substantially. In the experiments below, this mismatch remains visible in the studied industrial item catalog.

Representation compatibility has been studied through alignment and continuity of training. Orthogonal Low Rank Embedding Stabilization addresses version mixing through low-rank representations and orthogonal alignment~\cite{zielnicki2025stabilization}. TwHIN considers warm starts and historical regularization in a large-scale embedding workflow~\cite{elkishky2022twhin}, while GraphSAIL studies structure-preserving incremental recommendation~\cite{xu2020graphsail}. These studies motivate comparing post-processing, initialization, and inherited training states. Our downstream consumer is retrained for each evaluated configuration on mixed daily versions; the objective is compatibility within this training workflow.

Geometric continuity also needs to be assessed together with the representation's use. Prior work examines downstream instability of word embeddings~\cite{leszczynski2020instability} and the interpretation of embedding cosine similarity~\cite{steck2024cosine}. Spherical text embeddings provide precedent for restricting representation geometry~\cite{meng2019spherical}. Parameter rescaling in warm-started neural training offers another related perspective~\cite{ash2020warmstart}. Our global rescaling of both SGNS tables uses a reference mean item norm and no added perturbation; our spherical experiments constrain SGNS itself. We examine these interventions as existing or adapted components of an industrial pipeline.

We conduct systematic ablations and controlled comparisons across Word2Vec and an item--item graph adaptation of XSimGCL. Within each comparison, we retain the relevant data and evaluation protocol while adding, removing, or combining components: ID-based initialization and warm starts; OP, affine, MLP, and output-scale transformations; spherical constraints and temporal regularization; and downstream residual-noise or historical-version augmentation. Representation-family comparisons, fixed-version input strategies, encoder diagnostics, and deterministic-execution trials provide complementary evidence. The inventory contains 48 configurations, including 32 with CTR results under a common protocol.

Three questions organize the study:
\begin{enumerate}
\item Can upstream stabilization improve cross-day compatibility and CTR, and how do geometric and retrieval metrics help explain its effects?
\item Given the strong performance of warm starts, how should inherited scale accumulation be managed, and how does global state rescaling differ from per-item output normalization?
\item Can CTR learning further adapt to residual version differences, improving cross-day directional agreement in the encoder space?
\end{enumerate}

Our results support three corresponding findings. Multiple upstream interventions improve compatibility and GAUC, with joint evaluation revealing their trade-offs. Warm starts perform strongly, and their scale trajectories support global state management as a promising strategy. Finally, CTR-trained encoders improve directional agreement beyond dimension and architecture controls in multiple configurations. Results presents these findings first; Methods specifies the protocol and interventions, and Discussion interprets their mechanisms, practical implications, and scope.

\section{Results}\label{sec:results}
The common CTR protocol trains on Day 2--7 exposures, validates on Day 8, and tests on Day 9. Ordinary inputs use each exposure day's preceding embedding version. Main upstream geometry uses a fixed held-out cohort on Days 7--8; cosine measures same-item directional agreement and L2 retains the original scale. Co-occurrence Recall@100 retrieves next-day behavioral partners. Request-macro GAUC and LogLoss evaluate downstream ranking and probability prediction. The temporal and encoder analyses use their explicitly defined multi-day summaries (Section~\ref{sec:methods}). All reported metric results are relative percentage changes from the corresponding Word2Vec baseline (0\%); Section~\ref{sec:relative} defines the reporting convention.

\subsection{Multiple upstream interventions improve compatibility and CTR}\label{sec:stabilization}

\subsubsection{Base representations exhibit drift, and retrieval does not determine CTR ordering}
Independently trained Word2Vec versions define the cross-day geometry reference (0\%). XSimGCL has relative cosine and L2 changes of \M{xsimgcl_raw}{cosine} and \M{xsimgcl_raw}{l2}, respectively. Substantially overlapping training windows therefore do not ensure consistent coordinates for the same items.

XSimGCL has higher Recall@100 than Word2Vec (\M{xsimgcl_raw}{recall} versus \M{scratch_raw}{recall}), but lower GAUC (\M{xsimgcl_raw}{gauc} versus \M{scratch_raw}{gauc}). Across all four matched configurations in Supplementary Figure~S1c--d, XSimGCL has higher recall while Word2Vec has higher GAUC. Higher co-occurrence recall alone therefore does not justify replacing Word2Vec for this CTR task. The graph results refer throughout to the item--item adaptation described in Appendix~\ref{app:implementation}.

\begin{table}[htbp]\centering
\caption{Selected upstream configurations with ordinary DNN training. All metrics show relative changes (\%) from ordinary Word2Vec (0\%), rounded to one decimal. Geometry uses Days 7--8; recall uses next-day behavioral partners.}\label{tab:upstream}
\small\setlength{\tabcolsep}{4.5pt}\renewcommand{\arraystretch}{1.14}
\begin{tabular}{@{}lrrrrrr@{}}\toprule
Method & Cosine & L2 & Norm & Recall@100 & GAUC & LL\\\midrule
Word2Vec & 0.0 & 0.0 & 0.0 & 0.0 & 0.0 & 0.0 \\
W2V + OP & +93.4 & -48.2 & 0.0 & 0.0 & +2.6 & -0.2 \\
W2V + affine & +124.9 & -89.0 & -26.7 & -58.3 & +2.8 & -0.3 \\
W2V + MLP & +125.4 & -89.4 & -21.8 & -62.4 & +2.5 & -0.3 \\
ID W2V & +51.2 & -23.1 & 0.0 & -0.4 & +2.2 & \textbf{-0.6} \\
ID W2V + OP & +98.4 & -52.1 & 0.0 & -0.4 & +2.7 & -0.2 \\
ID W2V + unit + OP & +98.4 & -83.0 & -62.5 & -0.4 & +2.1 & -0.4 \\
Word2Vec + warm & +127.0 & -83.7 & +102.5 & -2.3 & +3.5 & -0.4 \\
Word2Vec + warm\_norm & +126.4 & -88.3 & +8.2 & +15.9 & \textbf{+4.0} & -0.5 \\
Word2Vec + warm\_unit & +127.0 & -97.2 & -62.5 & -2.3 & +3.1 & -0.5 \\
Word2Vec warm+spherical & +126.9 & -96.9 & -62.5 & +47.1 & +3.3 & -0.3 \\
XSimGCL & -102.4 & +31.2 & -0.9 & +76.4 & -0.5 & +0.2 \\
XSimGCL + OP & -35.0 & +8.1 & -0.9 & +76.4 & +2.2 & -0.2 \\
ID XSimGCL & +77.9 & -41.3 & -0.9 & +77.4 & +1.9 & -0.3 \\
ID XSimGCL + OP & +81.7 & -43.9 & -0.9 & +77.4 & +2.4 & -0.4 \\
ID XSimGCL + unit + OP & +81.7 & -78.2 & -62.5 & +77.4 & +2.2 & -0.3 \\
ID spherical W2V + OP & +116.8 & -89.4 & -62.5 & -3.9 & +1.8 & -0.2 \\
\bottomrule\end{tabular}
\smallnote{W2V: Word2Vec; ID: ID-init. Recall for OP and unit-normalization variants follows cosine-ranking invariance. Higher GAUC/recall and lower LogLoss are better; L2 and norm also depend on scale. Bold marks unrounded optima; rounding can produce ties.}
\end{table}

\subsubsection{Initialization control and alignment both provide positive evidence}
ID-init gives the same item the same initial vector across dates. For Word2Vec, ID-init raises cosine by \M{word2vec_idinit_raw}{cosine} and GAUC by \M{word2vec_idinit_raw}{gauc} relative to the ordinary baseline. For XSimGCL, relative cosine change increases from \M{xsimgcl_raw}{cosine} to \M{xsimgcl_idinit_raw}{cosine}, and relative GAUC change from \M{xsimgcl_raw}{gauc} to \M{xsimgcl_idinit_raw}{gauc}. Both pipelines show improved directional agreement together with better downstream prediction (Supplementary Figure~S1a,d).

Post-hoc OP alignment raises observed GAUC in all four comparisons formed by the two representation families and ordinary versus ID-init training. Word2Vec cosine and GAUC increase by \M{scratch_procrustes}{cosine} and \M{scratch_procrustes}{gauc}, respectively; ID-init Word2Vec with OP reaches a relative GAUC change of \M{word2vec_idinit_procrustes}{gauc}. A common orthogonal transform preserves within-version inner products, norms, and cosine retrieval rankings. These comparisons therefore show that cross-version alignment can benefit CTR while preserving the original retrieval relationships. The request-paired analysis supports a positive Word2Vec-to-OP GAUC difference on the evaluated test day (Table~\ref{tab:paired-bootstrap}, Appendix~\ref{app:statistics}).

The Word2Vec initialization comparisons in Supplementary Figure~S7a,d improve cosine and GAUC both with and without OP. These cumulative gains concern the complete workflows: provenance and parallel optimization also differ, precluding attribution solely to fixed initialization.

\subsubsection{More flexible matching has distinct geometric and retrieval effects}
Affine and MLP mappings yield relative cosine changes of \M{aligned_scratch_affine}{cosine} and \M{aligned_scratch_mlp}{cosine}, with Recall@100 changes of \M{aligned_scratch_affine}{recall} and \M{aligned_scratch_mlp}{recall}. Their relative GAUC changes, \M{aligned_scratch_affine}{gauc} and \M{aligned_scratch_mlp}{gauc}, lie on opposite sides of the OP result (Table~\ref{tab:upstream}; Supplementary Figure~S2a,c,d). Higher cosine therefore does not establish superiority over OP: recall is markedly lower and GAUC does not consistently improve.

Export-time unit normalization preserves cosine and cosine retrieval rankings while reducing raw L2 and removing item-specific norm information. Neither tested ID-init+OP pair improves GAUC after normalization; LogLoss improves only for Word2Vec (Supplementary Figure~S2a--d). Thus, shrinking L2 through output scaling does not itself establish a better CTR representation.

Unit-norm constraints during Word2Vec training do not ensure cross-day alignment: ID-init and OP remain useful within the spherical series. In the independently restarted series, ID-init spherical Word2Vec+OP has higher cosine but lower GAUC than unconstrained ID-init Word2Vec+OP (Table~\ref{tab:all}; Supplementary Figure~S8a,d).

\subsubsection{Geometric trends are informative alongside retrieval and CTR}
Across the 17 upstream configurations with ordinary DNN results, cosine has a positive descriptive association with GAUC and L2 a negative association (Supplementary Figure~S5a--b). Greater directional stability and smaller raw L2 tend to accompany higher GAUC, although shared artifacts, mixed scales, and different interventions preclude a universal causal interpretation.

The family separation in Supplementary Figure~S5c shows that recall is strongly associated with the representation family. Its mixed-configuration association with GAUC is weakly negative; this does not imply that improving recall harms CTR. Cosine and L2 can screen for reduced drift, but method selection still requires GAUC, LogLoss, and retention of behavioral retrieval relationships.

\FloatBarrier

\subsection{Warm starts perform strongly and motivate scale management}\label{sec:warm}

\subsubsection{Inherited states yield strong cross-day compatibility}
\begin{figure}[!htbp]\centering
\includegraphics[trim=0 0 0 28.8bp,clip,width=\textwidth]{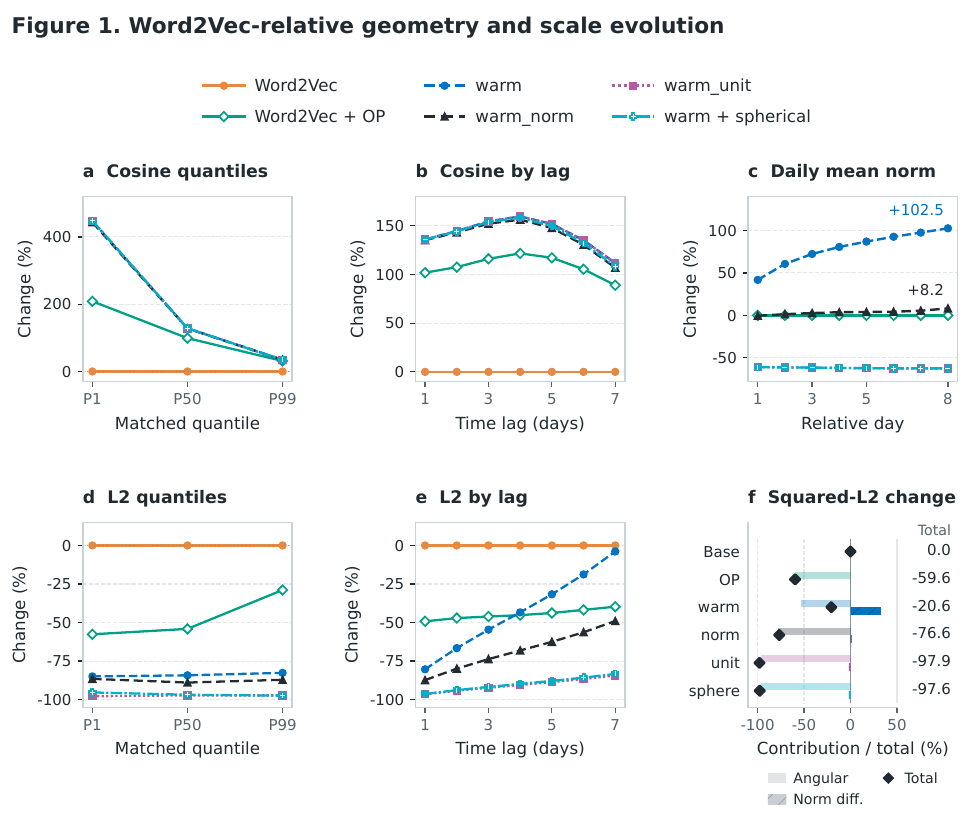}
\caption{Word2Vec-relative cross-day changes (\%) on a fixed held-out cohort. (a, d) Days 7--8 geometric quantiles compared with the corresponding Word2Vec quantile; lines connect the 1st, 50th, and 99th percentiles. (b, e) Cosine and L2 changes relative to Word2Vec at each lag. (c) Norm changes relative to Word2Vec on each day, including unit-constrained variants. (f) Day 1--8 angular and norm-difference contributions: each component difference from Word2Vec is divided by the Word2Vec squared-L2 total and multiplied by 100. Diamonds mark the total percentage change, equal to the sum of the two contributions. Base denotes Word2Vec; norm, unit, and sphere denote warm\_norm, warm\_unit, and warm+spherical. The reference is 0\% throughout.}\label{fig:warm-temporal}
\end{figure}

Among configurations with daily versions and ordinary DNN training, the Word2Vec warm family has the highest observed GAUC values (Table~\ref{tab:upstream}; Supplementary Figure~S7d). Compared with independent Word2Vec, warm raises adjacent-day cosine by \M{aligned_warm}{cosine} and GAUC by \M{aligned_warm}{gauc} (Figure~\ref{fig:warm-temporal}).

\FloatBarrier

\subsubsection{High directional agreement coexists with accumulated scale}
We track the same held-out item cohort over eight days to examine changes beyond adjacent-day cosine. Relative to Word2Vec on each day, warm mean-norm change increases from +41.8\% on Day 1 to +102.5\% on Day 8. Its L2 changes relative to Word2Vec at the corresponding lag are -80.3\% at a one-day lag and -3.9\% at a seven-day lag (Figure~\ref{fig:warm-temporal}). The former averages seven date pairs; the latter contains Day 1--8 only.

The squared-distance identity in Section~\ref{sec:diagnostics} separates the squared norm-difference and angular terms. At a seven-day lag, warm has a total squared-L2 change of -20.6\%. Figure~\ref{fig:warm-temporal}f shows how the angular and squared norm-difference changes contribute to this total, each divided by the Word2Vec total and multiplied by 100. This establishes a substantial geometric contribution from unequal norms over the observed interval. It is a decomposition of squared distance, rather than a causal allocation of training effects; the angular term also depends on norms.

\subsubsection{Global state rescaling attains the highest observed GAUC}
\WN\ limits relative norm growth: its day-matched norm change moves from -0.5\% to +8.2\%. Its one- and seven-day L2 changes are -87.3\% and -49.0\%. It attains the highest observed relative GAUC change among the 32 CTR configurations, \M{aligned_warm_norm}{gauc}, and a recall change of \M{aligned_warm_norm}{recall}, compared with warm recall change \M{aligned_warm}{recall}. \WU\ preserves warm's cosine relationships and recall, but has a GAUC change of \M{aligned_warm_unit}{gauc}; it has the lowest LogLoss among these three variants.

Word2Vec warm+spherical instead maintains unit item and context norms during training (Figure~\ref{fig:warm-temporal}). Its relative recall change reaches \M{warm_sphere}{recall}, exceeding \WN, but its relative GAUC change is lower at \M{warm_sphere}{gauc} and LogLoss is higher (Table~\ref{tab:upstream}; Supplementary Figure~S7). Thus, stronger retrieval under a strict norm constraint does not ensure better CTR. \WN\ remains the highest observed GAUC configuration. Its paired intervals against warm and \WU\ include zero (Table~\ref{tab:paired-bootstrap}); no paired interval was computed for warm+spherical.

\FloatBarrier

\subsection{CTR learning further alleviates residual drift in encoder space}\label{sec:encoder}

\subsubsection{Trained encoders achieve the highest directional agreement for most methods}
The CTR-trained encoder achieves the highest cross-day cosine among the examined representations for four of six upstream methods. The same four methods lead under adjacent-day, train-to-validation, and train-to-test aggregation. Adaptation thus extends to later embedding versions that were not used for gradient training, supporting further mitigation of residual directional drift in the encoder output.

We compare the original input with four lower-dimensional representations of the same size: a fixed random orthogonal projection, a train-only uncentered SVD projection, an untrained encoder, and the saved CTR-trained encoder. The untrained encoder has the same architecture and output dimension. Every mapping is shared across dates, and the same eligible items are evaluated for each representation within a date pair. Table~\ref{tab:encoder} reports train-to-test cosine. Complete cosine and L2 values for all three date scopes are reported in Tables~\ref{tab:encallcos} and~\ref{tab:encalll2}.

\begin{table}[htbp]\centering
\caption{Train-to-test relative cosine changes (\%). The original Word2Vec input is the shared baseline (0\%); bold marks the highest unrounded value within a row.}\label{tab:encoder}
\small\setlength{\tabcolsep}{4.5pt}\renewcommand{\arraystretch}{1.14}
\begin{tabular}{@{}lrrrrr@{}}\toprule
Method & Input & Random & SVD & Untrained & CTR\\\midrule
W2V & 0.0 & -12.9 & +40.1 & +72.4 & \textbf{+72.6} \\
W2V + OP & +104.4 & +101.5 & +127.4 & +120.6 & \textbf{+131.4} \\
ID W2V + OP & +109.3 & +112.5 & +128.9 & +126.0 & \textbf{+130.9} \\
ID XSimGCL + OP & +83.6 & +83.1 & +99.3 & +108.0 & \textbf{+118.2} \\
W2V + warm\_norm & +135.3 & +134.4 & \textbf{+140.7} & +137.4 & +137.1 \\
ID spherical W2V + OP & +130.2 & +130.0 & +137.2 & \textbf{+140.6} & +137.2 \\
\bottomrule\end{tabular}
\smallnote{Training versions are Days 1--6 and the test version is Day 8. A common denominator preserves comparisons across columns. Displayed ties can result from rounding. Full relative L2 changes appear in Appendix~\ref{app:encoder}.}
\end{table}

At the same output dimension, the CTR-trained Word2Vec+OP encoder exceeds the untrained and SVD controls in train-to-test cosine by 10.8 and 4.0 percentage points on the Word2Vec-relative change scale, respectively (Table~\ref{tab:encoder}). This provides evidence of additional adaptation through CTR learning beyond dimensional compression and network structure.

The gain is small for unaligned Word2Vec: 0.1 percentage points on the same scale above the untrained encoder. For \WN\ and ID-init spherical Word2Vec+OP, trained train-to-test cosine ranks third and second, respectively, within 3.6 percentage points of each method's maximum on that scale.

\subsubsection{Additional data augmentation has representation-dependent benefits}\label{sec:downstream}
Explicit augmentation has no consistent benefit across representations: strength-0.5 noise raises GAUC in three of seven pairs and lowers it in four, while both all-history comparisons lower GAUC. For Word2Vec+OP, the test-GAUC maximum at noise strength 0.5 differs from the validation-BCE preference for 1.0. Fixed training versions improve GAUC only for Word2Vec+OP among the three tested pairs; this strategy also changes information timing and coverage. These comparisons support representation-specific selection rather than default noise or history expansion. Table~\ref{tab:all} retains all results; Supplementary Figures~S3--S4 show the fixed-input and noise comparisons.

\FloatBarrier
\section{Methods}\label{sec:methods}
\subsection{Data, embedding versions, and temporal pairing}\label{sec:protocol}
We use proprietary data from LY Corporation's display-advertising operations: behavioral sequences for item embeddings, and logged ad impressions and clicks for CTR training and evaluation. Items are identified by a service-specific namespace and item identifier. The upstream pipeline produces daily item embeddings from recent behavioral data using overlapping windows. The experimental item catalogs contain at least one million items; inherited vocabularies can grow larger. Retrieval therefore uses a fixed common candidate set.

Days are indexed relative to the experimental window. Table~\ref{tab:split} specifies the ordinary feature join. The CTR data are extracts for research, rather than service-wide traffic totals. Training and testing each use at least one million logged examples. Two service-defined request categories are excluded consistently from evaluation. Labels refer to exposed items in the logs, not unexposed items elsewhere in the catalog.

\begin{table}[htbp]\centering
\caption{Temporal split and ordinary embedding-version mapping.}\label{tab:split}
\small\setlength{\tabcolsep}{4.5pt}\renewcommand{\arraystretch}{1.14}
\begin{tabular}{@{}llll@{}}\toprule
Split & Exposure dates & Embedding dates & Purpose\\\midrule
Training & Day 2--7 & Day 1--6, paired daily & Fit CTR model\\
Validation & Day 8 & Day 7 & Select checkpoint\\
Test & Day 9 & Day 8 & Final evaluation\\\bottomrule
\end{tabular}
\end{table}

\subsection{Systematic ablations and comparison design}
Component ablations add, remove, or combine operations within a base representation or DNN input protocol. Training-trajectory changes induced by a component are part of its effect; randomness and accompanying implementation changes are documented separately. Family comparisons, feature-version strategies, and engineering diagnostics complement these ablations (Table~\ref{tab:design}). The inventory is not a full factorial design, and unmeasured outcomes remain missing.

\begin{table}[htbp]\centering\small
\caption{Experimental groups, component changes, and evaluation questions.}\label{tab:design}
\setlength{\tabcolsep}{3.5pt}\renewcommand{\arraystretch}{1.14}
\begin{tabularx}{\textwidth}{@{}>{\raggedright\arraybackslash}p{106pt}>{\raggedright\arraybackslash}X>{\raggedright\arraybackslash}p{158pt}@{}}\toprule
Group & Baseline and intervention & Question\\\midrule
Representation family & Word2Vec versus item-graph XSimGCL & Retrieval and CTR ordering \\
Initialization & Base pipeline + ID-init & Consistent initial directions \\
Alignment & Same embeddings + OP & Retrieval-preserving compatibility \\
Mapping flexibility & No map, OP, affine, MLP & Alignment versus item relationships \\
Output scale & Same outputs + unit; fixed OP & Removal of per-item norms \\
Inheritance and scale & Independent training; warm; + unit/norm & Continuity and inherited scale \\
Spherical constraint & SGNS versus spherical; ID-init/OP/warm & Training-time norm restriction \\
Temporal regularization & ID-init + temporal; with/without OP & Historical direction constraints \\
Residual noise & Same DNN input + noise; strength sweep & Representation-specific benefit \\
Historical versions & No expansion; 50\% replacement; all history & Additional same-item versions \\
Training version & Daily pairing versus fixed Day 7 & Feature-preparation strategy \\
Determinism & One worker; synchronous pilots & Repeatability, quality, and cost \\
\bottomrule\end{tabularx}\end{table}

The basic operations are initialization (ID-init), inheritance (warm), post-processing (OP, affine, MLP, unit, norm), training constraints or regularization (spherical and temporal), and downstream augmentation (noise and historical versions). \WU\ is warm plus output unit normalization; \WN\ is warm plus daily global state rescaling. Norm is classified by its position after each day's training, although inheriting the rescaled state changes subsequent optimization. It is evaluated within \WN, not as an independently measured operation on every family.

OP fits a shared orthogonal mapping from unit-direction anchors and applies it to the full, potentially non-unit output. Unit+OP uses the saved directional transform; a unit-first refit was verified to give the same matrix. ID-init fixes the initial vector rather than freezing it or controlling every later update. The two-day temporal trials have no full CTR result; warm+spherical has a complete eight-day chain and ordinary DNN evaluation. The 48 configurations include reused upstream artifacts and small pilots; they are not independent replications. Appendix~\ref{app:implementation} summarizes the core interventions and comparison conditions.

\subsection{Inherited state and scale management}\label{sec:scale}
The inherited-training variants all use Word2Vec. Word2Vec + warm inherits both item and context states after the initial day. \WU\ unit-normalizes only exported item vectors from that same chain, leaving the state inherited by the next training day unchanged. \WN\ is a separate chain that rescales the entire item table $W_t$ and context table $C_t$ after each day:
\begin{equation}\label{eq:scale}
s_t=m_0/m_t,\qquad W_t\leftarrow s_tW_t,\qquad C_t\leftarrow s_tC_t,
\end{equation}
where $m_t$ is the full-vocabulary mean item norm and $m_0$ is the reference initialization mean. Each rescaling preserves current item directions and relative norms. Per-item unit normalization instead removes relative norm differences from the export. The full vocabulary evolves, so the fixed diagnostic cohort need not retain an exactly constant mean norm under \WN. Warm+spherical instead inherits unit-norm states and enforces the spherical SGNS constraint throughout training (Appendix~\ref{app:implementation}).

\subsection{Joint evaluation metrics}\label{sec:relative}
For the same item in versions $t$ and $s$, we report
\begin{equation}
c_{t,s}=\operatorname{mean}_i\frac{x_{i,t}^{\top}x_{i,s}}{\|x_{i,t}\|_2\|x_{i,s}\|_2},\qquad
d_{t,s}=\operatorname{mean}_i\|x_{i,t}-x_{i,s}\|_2.
\end{equation}
Cosine measures directional agreement; unnormalized L2 also depends on scale and dimension. Mean item norm is reported to aid interpretation. The main upstream comparison uses Days 7--8 and a fixed cohort of common evaluation items, disjoint from the transformation-fitting anchors. The norm column refers to Day 8.

Co-occurrence Recall@100 uses Day 8 vectors and exact cosine retrieval over a fixed common candidate set, excluding the query itself. Sampled queries are evaluated only when they have eligible Day 9 targets; exact candidate, query, and target counts are withheld. For eligible target set $T(q)$ and retrieved neighbors $N_{100}(q)$,
\begin{equation}
\mathrm{Recall@100}=\frac1{|Q|}\sum_{q\in Q}\frac{|N_{100}(q)\cap T(q)|}{|T(q)|}.
\end{equation}
Candidates, queries, and targets are fixed across main comparisons. Recall for OP and per-item unit-normalization variants is derived from the corresponding untransformed embeddings by cosine-ranking invariance; finite-precision tie-breaking is not claimed to be bitwise invariant. This behavioral recall is distinct from cross-version retrieval of the same item.

GAUC is the equal-request mean of within-request AUC:
\begin{equation}
\mathrm{GAUC}=\frac1{|G|}\sum_{g\in G}\mathrm{AUC}_g,\quad
G=\{g:g\text{ has both positive and negative labels}\}.
\end{equation}
Requests are the groups, rather than users, item namespaces, or the whole catalog. LogLoss averages binary prediction loss across all eligible test rows:
\begin{equation}
\mathrm{LogLoss}=-\frac1N\sum_{j=1}^{N}[y_j\log p_j+(1-y_j)\log(1-p_j)].
\end{equation}
The metrics respectively assess within-request discrimination and probability prediction. Both are offline measures.

\paragraph{Word2Vec-relative reporting.}
We use Word2Vec as the baseline and report all observed metric results as relative percentage increases or decreases from the corresponding baseline.

\subsection{CTR model and training}
We use a downstream DNN to test whether CTR learning can further reduce cross-day mismatch in item co-occurrence embeddings. Precomputed embeddings remain fixed during CTR training; a shared item encoder maps embeddings from all dates to a lower-dimensional space for the geometry analysis. Relevant comparisons retain the same labels, feature processing, and evaluation rules while varying the item representation or input strategy.

Checkpoints are selected using validation loss. The trained encoder is compared with an untrained encoder of the same architecture and with projections at the same output dimension. Encoder outputs are evaluated without additional normalization. These are chiefly single-run comparisons.

\subsection{Temporal trajectories and encoder diagnostics}\label{sec:diagnostics}
The temporal diagnostic uses the same held-out item cohort over eight dates and all 28 date pairs. A lag of $k$ days averages the $8-k$ available date-pair means equally. The exact per-item identity
\begin{equation}\label{eq:decomposition}
\|x-y\|_2^2=(\|x\|_2-\|y\|_2)^2+2\|x\|_2\|y\|_2(1-\cos(x,y))
\end{equation}
is averaged over items to separate the squared norm-difference and angular terms. It decomposes mean squared L2, not mean L2. The long-lag summaries have fewer available date pairs.

Encoder diagnostics reuse six saved upstream/CTR model pairs. Each date pair uses the same eligible full-catalog items across input, random projection, SVD, untrained encoder, and trained encoder representations. Each method uses one fixed mapping across all dates. SVD is fitted only on training-vector dates Day 1--6, with equal date weights and equal item weights within each date; there is no centering or whitening. The untrained network is a fresh fixed-seed initialization, not the original initialization checkpoint of the trained model.

Adjacent-day geometry averages seven date-pair means. Train-to-validation and train-to-test first average each item's available training dates, then give each eligible item equal weight. Validation and test vector dates are Day 7 and 8. The same item may occur during training even though its later vector version does not. All encoder outputs are evaluated in inference mode without additional unit normalization. The controls are summarized in Appendix~\ref{app:implementation}, and complete measurements in Appendix~\ref{app:encoder}.

\subsection{Paired uncertainty}
Paired GAUC comparisons use saved predictions on matched requests. We draw 20,000 paired bootstrap resamples of requests containing both labels and report percentile 95\% intervals. These intervals are conditional on the saved models and one test day; Appendix~\ref{app:statistics} reports the comparisons.

\section{Discussion}
\subsection{Joint evaluation identifies useful stabilization}
The comparisons support cross-day stabilization as a practical way to improve multi-version CTR inputs. Useful interventions must also retain task-relevant item relationships: stronger geometric matching can accompany retrieval loss or lower GAUC. Geometry diagnoses drift, while retrieval and CTR determine whether the changed representation remains useful.

\subsection{Global scale management and inherited optimization}\label{sec:discussion-scale}
The combination of controlled norm growth and the highest observed GAUC favors managing the inherited training state while retaining relative item norms at each rescaling. Scaling both SGNS tables multiplies their inner products by $s_t^2$, changing subsequent logits and gradients. The relative recall change from \M{aligned_warm}{recall} to \M{aligned_warm_norm}{recall} cannot arise from a final common output rescaling alone, because that preserves cosine rankings. Scale management therefore acts through the continuing training process, with coupled effects on scale and optimization. Warm+spherical further raises recall but does not surpass \WN\ in GAUC or LogLoss. Training-time unit norms therefore do not suffice for the best observed CTR; the comparison also changes the loss, optimization settings, and initial contexts.

\subsection{CTR learning provides a further adaptation stage}
CTR learning can further alleviate residual directional drift in encoder space. The upstream vectors remain fixed, while the encoder learns a mapping shared across dates. Gains against same-dimensional controls support an additional contribution from task learning. This learned adaptation complements upstream stabilization.

Learning is assessed primarily through cosine under same-dimensional controls. Raw L2 is auxiliary: an unscaled random orthonormal projection to a lower-dimensional space can reduce the expected squared length of a fixed difference vector even without learned alignment.

\subsection{Practical recommendations and online evaluation}\label{sec:engineering}
For an existing daily Word2Vec pipeline that prioritizes limited code changes and low operational overhead, OP is a practical initial baseline. It preserves the training procedure and within-version cosine retrieval relationships, requiring an added alignment step and reference-version management. Where additional state management is feasible, \WN\ can be evaluated against this baseline through online A/B testing to determine whether its offline gains translate into online benefits that justify the cost of retaining and recovering both embedding tables and the vocabulary across days. Execution-repeatability and parallelization diagnostics are reported in Appendix~\ref{app:determinism} and Supplementary Figure~S6.

\subsection{Scope and limitations}\label{sec:limitations}
The paired intervals support a positive Word2Vec-to-OP GAUC difference conditional on the saved models and test day; \WN\ versus warm and \WU\ intervals include zero (Table~\ref{tab:paired-bootstrap}). These exploratory, pointwise intervals have no multiple-comparison or selection adjustment.

The study covers one industrial setting and one main window, chiefly with one training run per configuration; later choices were informed by test results. Warm chains are independently trained and asynchronous, family comparisons differ in construction and objectives, and spherical training changes several optimization settings. Warm's scale, relative-norm, and optimization effects are not isolated. Some historical scratch-input ordering hashes are unavailable. GAUC and recall use restricted eligible-request and query populations; corpus size does not replace independent seeds or periods. The bootstrap omits retraining variability and within-user request dependence.

The untrained control uses a fresh initialization, preventing precise causal attribution to dimension, architecture, or supervision. We did not test how much geometry mediates GAUC, or retrain CTR on SVD inputs. Encoder rankings are descriptive point estimates.

All predictive results are offline; no online CTR lift is reported.

\section{Conclusion}
Daily item co-occurrence embeddings introduce cross-version compatibility challenges in multi-day CTR learning. Systematic ablations and comparisons across two representation pipelines show that initialization control, post-hoc alignment, and warm starts can improve compatibility and downstream prediction. Joint evaluation of geometry, behavioral retrieval, and CTR identifies the benefits and trade-offs of these interventions.

Warm starts perform strongly and exhibit accumulated scale beyond their high directional agreement. Global state rescaling limits the observed norm growth and attains the highest observed GAUC, supporting its prioritization in the studied inherited-training workflow and its separate assessment from per-item output normalization.

CTR learning can further alleviate residual directional drift in encoder space: trained encoders achieve the highest cross-day cosine for four of six upstream methods. These findings support complementary roles for upstream stabilization and downstream task learning, with practical choices informed by both joint evaluation and the cost of maintaining daily states and versions.

\clearpage
\appendix
\section*{Appendix}\vspace{-0.75\baselineskip}

\section{Complete Experimental Inventory}\label{app:inventory}
The inventory covers 48 configurations and 32 CTR results; \NA\ denotes missing measurements. OP/unit recall follows cosine-ranking invariance. Superscripts mark reused upstream metrics (\rr), two-day temporal pilots (\pp), and small parallel-training pilots (\ssm). Otherwise, geometry uses Day 7--8 and norm uses Day 8. W2V means Word2Vec; ID means ID-init. All observed results use the relative percentage changes defined in Section~\ref{sec:relative}; bold CTR values identify the best unrounded point estimates. Small parallel pilots use their own matched asynchronous Word2Vec references and are not comparable to the main evaluation.

\begingroup\small\setlength{\tabcolsep}{4.5pt}\renewcommand{\arraystretch}{1.14}
\begin{longtable}{@{}r l rrrrrr@{}}
\caption{Complete 48-configuration inventory in relative changes (\%) from Word2Vec (0\%). Pilot rows use matched pilot baselines. Scopes and recall derivations are described in Appendix~\ref{app:inventory}.}\label{tab:all}\\
\toprule ID & Configuration & Cosine & L2 & Norm & Recall & GAUC & LL\\\midrule\endfirsthead
\multicolumn{8}{l}{\textit{Table \thetable\ continued}}\\\toprule
ID & Configuration & Cosine & L2 & Norm & Recall & GAUC & LL\\\midrule\endhead
\midrule\multicolumn{8}{r}{\textit{Continued on next page}}\\\endfoot
\bottomrule\endlastfoot
\multicolumn{8}{@{}l}{\textbf{A. Representation families}}\\*
1 & Word2Vec & 0.0 & 0.0 & 0.0 & 0.0 & 0.0 & 0.0 \\
2 & XSimGCL & -102.4 & +31.2 & -0.9 & +76.4 & -0.5 & +0.2 \\
\multicolumn{8}{@{}l}{\textbf{B. Post-processing}}\\*
3 & W2V + OP & +93.4 & -48.2 & 0.0 & 0.0 & +2.6 & -0.2 \\
4 & W2V + unit & 0.0 & -63.2 & -62.5 & 0.0 & \NA & \NA \\
5 & W2V + unit + OP & +93.4 & -81.6 & -62.5 & 0.0 & \NA & \NA \\
6 & W2V + affine & +124.9 & -89.0 & -26.7 & -58.3 & +2.8 & -0.3 \\
7 & W2V + MLP & +125.4 & -89.4 & -21.8 & -62.4 & +2.5 & -0.3 \\
8 & XSimGCL + OP & -35.0 & +8.1 & -0.9 & +76.4 & +2.2 & -0.2 \\
9 & XSimGCL + unit & -102.4 & -50.4 & -62.5 & +76.4 & \NA & \NA \\
10 & XSimGCL + unit + OP & -35.0 & -58.7 & -62.5 & +76.4 & \NA & \NA \\
\multicolumn{8}{@{}l}{\textbf{C. Initialization and warm starts}}\\*
11 & ID W2V & +51.2 & -23.1 & 0.0 & -0.4 & +2.2 & \textbf{-0.6} \\
12 & ID W2V + OP & +98.4 & -52.1 & 0.0 & -0.4 & +2.7 & -0.2 \\
13 & ID W2V + unit & +51.2 & -71.5 & -62.5 & -0.4 & \NA & \NA \\
14 & ID W2V + unit + OP & +98.4 & -83.0 & -62.5 & -0.4 & +2.1 & -0.4 \\
15 & Word2Vec + warm & +127.0 & -83.7 & +102.5 & -2.3 & +3.5 & -0.4 \\
16 & Word2Vec + warm\_norm & +126.4 & -88.3 & +8.2 & +15.9 & \textbf{+4.0} & -0.5 \\
17 & Word2Vec + warm\_unit & +127.0 & -97.2 & -62.5 & -2.3 & +3.1 & -0.5 \\
18 & Word2Vec warm+spherical & +126.9 & -96.9 & -62.5 & +47.1 & +3.3 & -0.3 \\
19 & ID XSimGCL & +77.9 & -41.3 & -0.9 & +77.4 & +1.9 & -0.3 \\
20 & ID XSimGCL + OP & +81.7 & -43.9 & -0.9 & +77.4 & +2.4 & -0.4 \\
21 & ID XSimGCL + unit & +77.9 & -77.3 & -62.5 & +77.4 & \NA & \NA \\
22 & ID XSimGCL + unit + OP & +81.7 & -78.2 & -62.5 & +77.4 & +2.2 & -0.3 \\
\multicolumn{8}{@{}l}{\textbf{D. Losses and constraints}}\\*
23 & Spherical W2V & -61.4 & -55.0 & -62.5 & -6.3 & \NA & \NA \\
24 & Spherical W2V + OP & -14.0 & -61.6 & -62.5 & -6.3 & \NA & \NA \\
25 & ID spherical W2V & +97.8 & -82.4 & -62.5 & -3.9 & \NA & \NA \\
26 & ID spherical W2V + OP & +116.8 & -89.4 & -62.5 & -3.9 & +1.8 & -0.2 \\
27 & ID W2V + temporal\pp & +88.4 & -44.7 & \NA & +3.8 & \NA & \NA \\
28 & ID W2V + temporal + OP\pp & +110.9 & -63.5 & \NA & +3.8 & \NA & \NA \\
29 & ID XSimGCL + temporal\pp & +81.2 & -44.5 & \NA & +77.3 & \NA & \NA \\
30 & ID XSimGCL + temporal + OP\pp & +84.5 & -46.7 & \NA & +77.3 & \NA & \NA \\
\multicolumn{8}{@{}l}{\textbf{E. Fixed training input}}\\*
31 & W2V / fixed Day 7\rr & 0.0 & 0.0 & \NA & 0.0 & -0.2 & +0.4 \\
32 & W2V + OP / fixed Day 7\rr & +93.4 & -48.2 & \NA & 0.0 & +3.9 & -0.3 \\
33 & XSimGCL + OP / fixed Day 7\rr & -35.0 & +8.1 & \NA & +76.4 & +1.2 & -0.2 \\
\multicolumn{8}{@{}l}{\textbf{F. Downstream augmentation}}\\*
34 & W2V + OP / noise 0.25\rr & +93.4 & -48.2 & 0.0 & 0.0 & +3.0 & -0.2 \\
35 & W2V + OP / noise 0.5\rr & +93.4 & -48.2 & 0.0 & 0.0 & +3.3 & -0.3 \\
36 & W2V + OP / noise 1.0\rr & +93.4 & -48.2 & 0.0 & 0.0 & +2.9 & -0.3 \\
37 & W2V + OP / history 50\%\rr & +93.4 & -48.2 & 0.0 & 0.0 & +2.6 & -0.2 \\
38 & W2V + OP / all history\rr & +93.4 & -48.2 & 0.0 & 0.0 & +2.0 & -0.3 \\
39 & Word2Vec + warm\_norm / noise 0.5\rr & +126.4 & -88.3 & +8.2 & +15.9 & +3.8 & -0.5 \\
40 & W2V / all history\rr & 0.0 & 0.0 & 0.0 & 0.0 & 0.0 & +0.1 \\
41 & ID W2V + OP / noise 0.5\rr & +98.4 & -52.1 & 0.0 & -0.4 & +2.8 & -0.3 \\
42 & ID XSimGCL + OP / noise 0.5\rr & +81.7 & -43.9 & -0.9 & +77.4 & +2.2 & -0.4 \\
43 & ID W2V + unit + OP / noise 0.5\rr & +98.4 & -83.0 & -62.5 & -0.4 & +1.5 & -0.4 \\
44 & ID XSimGCL + unit + OP / noise 0.5\rr & +81.7 & -78.2 & -62.5 & +77.4 & +2.0 & -0.3 \\
45 & ID spherical W2V + OP / noise 0.5\rr & +116.8 & -89.4 & -62.5 & -3.9 & +2.3 & -0.3 \\
\multicolumn{8}{@{}l}{\textbf{G. Repeatability and parallelization}}\\*
46 & W2V / one worker & \NA & \NA & \NA & \NA & \NA & \NA \\
47 & Synchronous W2V / prototype 1\ssm & \NA & \NA & \NA & -84.9 & \NA & \NA \\
48 & Synchronous W2V / prototype 2\ssm & \NA & \NA & \NA & -58.4 & \NA & \NA \\
\end{longtable}
\endgroup

\subsection{Timing scope}
Supplementary Tables S1--S2 give comparison coverage and all 48 timing records. Embedding times include recorded I/O; OP rows reuse base times, excluding unmeasured OP/unit overhead. Word2Vec uses a separate two-run timing reference; graph means cover seven days, ID-init Word2Vec and warm chains eight days, independently restarted spherical Day 8, and temporal pilots two days. DNN times include validation where recorded. Shared artifacts incur no additional training; differing epochs, I/O, and hardware preclude controlled throughput comparisons.

\section{Core Experimental Interventions}\label{app:implementation}
This appendix summarizes the interventions needed to interpret the comparisons. Routine optimizer, architecture, and execution settings are omitted.

\subsection{Base representations}
Word2Vec uses SGNS on item co-occurrence sequences~\cite{mikolov2013distributed,barkan2016item2vec}; we export the item vectors. The graph baseline is a study-specific weighted item--item adaptation of XSimGCL~\cite{yu2022xsimgcl}, combining neighborhood aggregation and contrastive learning with embedding perturbations. Its construction and objective differ from the original recommendation model. A common positive rescaling sets its exported median norm to a fixed reference scale without individually unit-normalizing the vectors.

\subsection{Initialization and inherited states}
ID-init assigns each item a deterministic initial state from its namespace and item key, independently of date or vocabulary order. This controls initialization; vectors remain trainable and parallel update order is not fixed.

Warm starts from saved prior-day item vectors and reinitializes missing context states. Subsequent dates inherit both item and context tables while updating the vocabulary. \WU\ changes only the exported vectors from that warm chain; \WN\ is a separate chain that inherits globally rescaled states (Section~\ref{sec:scale}).

\subsection{Alignment and scale management}
OP fits an orthogonal map between current and preceding processed versions using anchor directions, then applies it to the full vectors. It preserves within-version inner products, norms, and cosine rankings. Affine and MLP alternatives instead fit unconstrained linear-with-intercept and nonlinear maps to the preceding processed vectors by squared-error matching. The first day is unchanged, and fitting anchors are disjoint from the evaluation cohort.

Output unit normalization removes per-item norms after training. The \WN\ operation in Section~\ref{sec:scale} rescales both inherited tables by a common factor, retaining relative item norms and affecting subsequent training.

\subsection{Spherical training constraint}
The spherical SGNS variant constrains both item and context vectors to unit norm during training~\cite{meng2019spherical}. This changes the optimization, unlike normalizing exported vectors afterward. Ordinary and ID-init variants differ in initialization; OP is applied after training where specified. Warm+spherical initializes from normalized prior-day item directions and fresh unit-norm contexts, then inherits both tables across days. Its loss, optimization settings, and initial contexts differ from ordinary warm, so the comparison does not isolate one operation.

\subsection{Temporal regularization}
Both temporal pilots restart with ID-init and encourage agreement with frozen preceding-day directions for common items; new items have no historical term. The graph variant adds a cosine-distance penalty. The Word2Vec variant alternates ordinary SGNS updates with directional blending toward the previous version while preserving current norms. These use different optimization procedures and are evaluated only in the short pilot scope identified in the inventory.

\subsection{Downstream augmentation and version selection}
Residual noise uses normalized differences between adjacent training-date embeddings:
\[
\delta_{i,t}=\frac{x_{i,t}-x_{i,t-1}}{(\|x_{i,t}\|_2+\|x_{i,t-1}\|_2)/2},
\qquad \widetilde x=x+\alpha\|x\|_2\epsilon\delta,
\quad \epsilon\in\{-1,+1\}.
\]
For a subset of eligible training inputs, residuals are sampled across items from the same method's training-period reservoir with random sign. Noise strengths are identified in the inventory. Validation/test inputs are unperturbed, and noisy inputs are not renormalized.

Historical augmentation either substitutes an eligible earlier version for selected rows or expands each row across available versions, retaining labels and other features. Partial replacement retains one input per row and falls back to the original when history is unavailable. Fixed-version training instead uses the designated later embedding version for earlier exposures; it therefore changes information timing and coverage. Validation/test mappings remain unchanged.

\subsection{Repeatability scope}\label{app:determinism}
Single-worker Word2Vec and ID-init XSimGCL were bitwise repeatable in the tested environment. Parallel Word2Vec variants retain scheduling variability. Synchronous prototypes control update ordering but also change sampling and optimization; they are separate implementations rather than isolated scheduling changes. Their repeatability was verified only in small pilots, using a separate same-day retrieval protocol. Supplementary Figure~S6 reports the relative runtime and retrieval results.

\subsection{Dimension and encoder controls}
Random orthogonal projection and uncentered SVD provide fixed mappings to the same output dimension as the encoder. Random projection has no additional rescaling. SVD uses only training dates, with equal date weights and equal item weights within each date; it uses neither labels nor held-out dates and applies no whitening.

The untrained encoder uses a fresh fixed initialization of the trained encoder's architecture. The trained encoder uses the validation-selected checkpoint. All mappings are shared across dates and evaluated without additional output normalization, allowing task learning to be compared with dimensional and architectural controls.

\subsection{Warm-start temporal diagnostics}\label{app:temporal}
Section~\ref{sec:diagnostics} defines the fixed-cohort lag averages and squared-distance decomposition in Figure~\ref{fig:warm-temporal}. \WN's scale target uses the full evolving vocabulary, distinct from the fixed cohort.
\FloatBarrier

\section{Complete Encoder Controls}\label{app:encoder}
Tables~\ref{tab:encallcos}--\ref{tab:encalll2} retain all 180 geometry values: six methods, five representations, three date scopes, and two metrics. Section~\ref{sec:diagnostics} defines aggregation. All transformed outputs share the same output dimension and are evaluated without extra normalization.

\begin{table}[htbp]\centering
\caption{Complete cosine changes (\%). Each scope uses its original Word2Vec input as a common baseline (0\%) across all columns and methods.}\label{tab:encallcos}
\small\setlength{\tabcolsep}{4.5pt}\renewcommand{\arraystretch}{1.14}
\begin{tabular}{@{}llrrrrr@{}}\toprule
Method & Scope & Input & Random & SVD & Untrained & CTR\\\midrule
W2V & Adjacent & 0.0 & -7.6 & +29.6 & +61.0 & +80.3 \\
 & Train--val & 0.0 & -10.5 & +35.0 & +67.5 & +68.9 \\
 & Train--test & 0.0 & -12.9 & +40.1 & +72.4 & +72.6 \\
\addlinespace[4pt]
W2V + OP & Adjacent & +96.2 & +94.3 & +113.2 & +108.2 & +117.7 \\
 & Train--val & +95.3 & +92.5 & +116.1 & +110.0 & +120.4 \\
 & Train--test & +104.4 & +101.5 & +127.4 & +120.6 & +131.4 \\
\addlinespace[4pt]
W2V + warm\_norm & Adjacent & +127.6 & +127.5 & +128.3 & +127.8 & +127.7 \\
 & Train--val & +125.6 & +124.9 & +129.5 & +127.1 & +127.0 \\
 & Train--test & +135.3 & +134.4 & +140.7 & +137.4 & +137.1 \\
\addlinespace[4pt]
ID W2V + OP & Adjacent & +101.3 & +103.6 & +114.8 & +113.3 & +117.7 \\
 & Train--val & +100.3 & +103.1 & +117.6 & +115.4 & +119.8 \\
 & Train--test & +109.3 & +112.5 & +128.9 & +126.0 & +130.9 \\
\addlinespace[4pt]
ID XSimGCL + OP & Adjacent & +85.0 & +84.5 & +96.3 & +100.2 & +108.7 \\
 & Train--val & +80.3 & +79.7 & +93.9 & +99.9 & +109.3 \\
 & Train--test & +83.6 & +83.1 & +99.3 & +108.0 & +118.2 \\
\addlinespace[4pt]
ID spherical W2V + OP & Adjacent & +119.1 & +119.1 & +123.3 & +124.9 & +123.4 \\
 & Train--val & +120.1 & +120.1 & +125.8 & +128.7 & +125.6 \\
 & Train--test & +130.2 & +130.0 & +137.2 & +140.6 & +137.2 \\
\addlinespace[4pt]
\bottomrule\end{tabular}

\end{table}
\begin{table}[htbp]\centering
\caption{Complete L2 changes (\%). Each scope uses its original Word2Vec input as a common baseline (0\%) across all columns and methods.}\label{tab:encalll2}
\small\setlength{\tabcolsep}{4.5pt}\renewcommand{\arraystretch}{1.14}
\begin{tabular}{@{}llrrrrr@{}}\toprule
Method & Scope & Input & Random & SVD & Untrained & CTR\\\midrule
W2V & Adjacent & 0.0 & -43.9 & -24.8 & -39.0 & -63.4 \\
 & Train--val & 0.0 & -44.3 & -27.8 & -39.7 & -59.0 \\
 & Train--test & 0.0 & -43.3 & -29.3 & -40.8 & -57.5 \\
\addlinespace[4pt]
W2V + OP & Adjacent & -49.2 & -71.3 & -67.7 & -67.2 & -79.3 \\
 & Train--val & -45.0 & -68.9 & -65.2 & -64.8 & -77.2 \\
 & Train--test & -44.6 & -68.7 & -65.1 & -64.7 & -76.5 \\
\addlinespace[4pt]
W2V + warm\_norm & Adjacent & -86.4 & -92.4 & -90.5 & -91.2 & -91.9 \\
 & Train--val & -72.0 & -84.2 & -80.9 & -82.1 & -83.9 \\
 & Train--test & -67.4 & -81.7 & -77.8 & -79.4 & -81.4 \\
\addlinespace[4pt]
ID W2V + OP & Adjacent & -53.2 & -73.5 & -69.4 & -70.1 & -79.6 \\
 & Train--val & -48.6 & -70.8 & -66.7 & -67.4 & -77.0 \\
 & Train--test & -47.8 & -70.6 & -66.4 & -66.8 & -76.6 \\
\addlinespace[4pt]
ID XSimGCL + OP & Adjacent & -43.8 & -68.4 & -68.9 & -60.9 & -76.0 \\
 & Train--val & -38.2 & -65.3 & -65.6 & -57.7 & -73.6 \\
 & Train--test & -35.8 & -64.0 & -64.2 & -56.9 & -72.8 \\
\addlinespace[4pt]
ID spherical W2V + OP & Adjacent & -89.1 & -93.8 & -93.9 & -84.9 & -85.0 \\
 & Train--val & -87.1 & -92.8 & -92.7 & -82.4 & -81.6 \\
 & Train--test & -86.5 & -92.4 & -92.4 & -81.7 & -81.2 \\
\addlinespace[4pt]
\bottomrule\end{tabular}

\end{table}\FloatBarrier

\FloatBarrier

\section{Paired Statistical Analysis}\label{app:statistics}
\begin{table}[htbp]\centering\small
\caption{Differences in Word2Vec-relative GAUC change (percentage points), with pointwise 95\% paired-bootstrap intervals. Positive differences favor the method on the right.}\label{tab:paired-bootstrap}
\setlength{\tabcolsep}{4.5pt}\renewcommand{\arraystretch}{1.14}
\begin{tabular}{@{}lrr@{}}\toprule
Comparison & $\Delta$ change (pp) & 95\% interval\\\midrule
Word2Vec $\rightarrow$ Word2Vec+OP & +2.6 & [+0.8, +4.3] \\
warm $\rightarrow$ \WN & +0.5 & [-0.6, +1.6] \\
warm $\rightarrow$ \WU & -0.4 & [-1.5, +0.7] \\
\WU $\rightarrow$ \WN & +0.9 & [-0.2, +2.1] \\
Word2Vec+OP $\rightarrow$ \WN & +1.5 & [+0.1, +2.9] \\
\bottomrule\end{tabular}\end{table}

Each bootstrap resample uses the same requests for both compared models. Table~\ref{tab:paired-bootstrap} reports differences on the Word2Vec-relative GAUC scale, using a fixed observed baseline. These exploratory intervals do not cover retraining or other test dates and have no multiple-comparison or selection adjustment.
\FloatBarrier
\end{document}